\documentclass[prd,
twocolumn,superscriptaddress,preprintnumbers,nofootinbib]{revtex4}
\usepackage{graphicx}
\usepackage{epsfig}
\usepackage{bm}
\usepackage{latexsym,amssymb,amsmath,amssymb,wasysym,float}

\usepackage{color}

\usepackage{scrextend}
\usepackage{mathrsfs}
\usepackage{enumitem}

\newcommand{\postscript}[2]{\setlength{\epsfxsize}{#2\hsize}
   \centerline{\epsfbox{#1}}}

\usepackage[usenames,dvipsnames]{xcolor}
\definecolor{orange}{cmyk}{0,0.5,1,0}
\definecolor{rossoCP3}{cmyk}{0,.88,.77,.40}
\definecolor{graa}{rgb}{0.8,0.8,0.8}
\definecolor{blaa}{rgb}{0.2,0.2,0.6}

\begin{document}

\preprint{MPP-2025-5}
\preprint{LMU-ASC 02/25}

\title{\color{rossoCP3} Two Micron-Size  Dark Dimensions}

\author{\bf Luis A. Anchordoqui}

\affiliation{Department of Physics and Astronomy,  Lehman College, City University of
  New York, NY 10468, USA
}

\affiliation{Department of Physics, 
 Graduate Center,  City University of
  New York,  NY 10016, USA
}

\affiliation{Department of Astrophysics,
 American Museum of Natural History, NY
 10024, USA
}

\author{\bf Ignatios Antoniadis}

\affiliation{High Energy Physics Research Unit, Faculty of Science, Chulalongkorn University, Bangkok 1030, Thailand}

\affiliation{Laboratoire de Physique Th\'eorique et Hautes \'Energies
  - LPTHE 
Sorbonne Universit\'e, CNRS, 4 Place Jussieu, 75005 Paris, France
}

\author{\bf Dieter\nolinebreak~L\"ust}

\affiliation{Max--Planck--Institut f\"ur Physik,  
 Werner--Heisenberg--Institut,
80805 M\"unchen, Germany
}

\affiliation{Arnold Sommerfeld Center for Theoretical Physics, 
Ludwig-Maximilians-Universit\"at M\"unchen,
80333 M\"unchen, Germany
}

\begin{abstract}
  \noindent Two extra dimensions of micron scale might
  simultaneously address the gauge and cosmological hierarchy
  problems. In our paper we examine various observational bounds in  scenarios with one and two large extra dimensions, to see if they are compatible with the micron scale. 
  We show that consistency with astrophysical 
  observations requires that two extra dimensions of micron scale must not admit
  isometries, whereby conservation of the extra dimensional momentum is
  violated, allowing the massive Kaluza-Klein modes
  of the graviton to decay to other lighter graviton
  modes. However, to remain consistent with cosmological observations
  two extra dimensions of micron scale require a delicately fine
  tuning of the temperature at which the universe enters the radiation
  dominated epoch. Diving into this fine-tuned scenario we also
 show that primordial black holes with masses in the
  range $10^8 \alt M_{\rm BH}/{\rm g} \alt 10^{21}$ could make all cosmological dark matter. 
\end{abstract}
\date{January 2025}
\maketitle

\section{General Idea}

Extra dimensions are in general thought to be compact, since a
four-dimensional (4D) description fits the world we perceive and
measure (so far) very well. It is this compactness that sets a new
length scale. A particularly attractive framework has as its main
feature large extra dimensions along which only gravity propagates, 
because Standard Model (SM) fields
are localized on a brane extended in the non-compact
dimensions~\cite{Arkani-Hamed:1998jmv}.
Most notably, this construct can be
embedded into string theory~\cite{Antoniadis:1998ig}. Within this framework the usual 4D graviton is complemented by a tower of Kaluza-Klein (KK) modes,
corresponding to the new available phase space in the bulk.
In the canonical example,
spacetime is a direct product of a non-compact 4D spacetime manifold
and a flat spatial $n$-torus of common linear size $2\pi R_{\rm KK}$ and
volume $V_n = (2\pi R_{\rm KK})^n$. The $(4+n)$ dimensional Planck scale
(or more generally the species scale $\Lambda_{\rm sp}$ where gravity becomes strong~\cite{Dvali:2007hz,Dvali:2007wp,Dvali:2009ks,Dvali:2010vm,Dvali:2012uq}) 
is related to the 4D reduced
Planck mass $M_p$ according to
\begin{equation}
  \Lambda_{\rm sp} = M_p^{2/(2+n)} \ V_n^{1/(2+n)} \, .
\end{equation}  

Besides, by combining this idea with conjectures of the Swampland
program~\cite{Vafa:2005ui,Palti:2019pca,vanBeest:2021lhn,Agmon:2022thq} a deep interplay between physics in the UV and the IR becomes exposed.
For example, the distance conjecture states that infinite
distance limits $\Delta \phi \to \infty$ in the moduli space of
massless scalar fields are accompanied by an infinite tower of
exponentially light states $m_{\rm KK} \sim e^{-\alpha \Delta \phi}$, where
distance and masses are measured in Planck
units~\cite{Ooguri:2006in}. Connected to the distance conjecture is
the anti-de Sitter (AdS) distance conjecture, which correlates the dark energy density
to the mass scale $m_{\rm KK}$ characterizing the infinite tower of
states,
\begin{equation}
m_{\rm KK}
\sim |\Lambda|^\alpha \,,
\label{AdSDC}
\end{equation}
as the negative AdS vacuum energy $\Lambda \to
0$, where $\alpha$ is an ${\cal O}(1)$ positive
constant~\cite{Lust:2019zwm}. 
Besides, under the hypothesis that
this scaling behaviour (\ref{AdSDC}) holds in de Sitter
(or quasi de Sitter) space, an unbounded number of massless modes should also
emerge
in the limit $\Lambda \to 0$.

Note that the KK (or string)
tower contains massive spin-2 bosons, and thus for positive $\Lambda \to 0$ the Higuchi
  bound~\cite{Higuchi:1986py} places an absolute upper limit to the
  exponent: $\alpha =1/2$. In addition, explicit string calculations
of the vacuum energy
set a lower bound: $\alpha =1/4$~\cite{Montero:2022prj} (see
e.g. also~\cite{Itoyama:1986ei,Itoyama:1987rc,Antoniadis:1991kh,Bonnefoy:2018tcp,Burgess:2023pnk,Branchina:2023ogv,Anchordoqui:2023laz,Branchina:2024ljd,Aoufia:2024awo,Basile:2024lcz}).
Now, the combination of the theoretically accessible parameter range $1/4\leq\alpha\leq1/2$
with the experimental constraints on deviations from Newton's
gravitational inverse-square law~\cite{Lee:2020zjt}, namely
\begin{equation}R_{\rm KK} \alt 30~\mu{\rm m}\, ,
\end{equation}
selects $\alpha=1/4$ as the preferred value and 
gives in this way a procedure to 
connect $R_{\rm KK}$ to the
dark energy scale $\Lambda/M_p^{4} \sim 10^{-120}$ (i.e. $\Lambda^{1/4} \sim 2.39~{\rm meV}$~\cite{Planck:2018vyg})
 as follows~\cite{Montero:2022prj}
\begin{equation}
  R_{\rm KK} \sim \lambda \ \Lambda^{-1/4}\simeq {\cal O}(\mu{\rm m}) \,,
\label{RperpLambda}  
\end{equation}  
where the proportionality factor is estimated to be within the range
$10^{-1} < \lambda < 10^{-4}$.

Actually, the relation (\ref{RperpLambda}) is  independent of the number $n$ of dark dimensions of micron size.
Taking into account the supernovae and neutron star constraints (see the discussion in the next section), the authors of~\cite{Montero:2022prj}
have argued in favor of a single  dark dimension  in the micron range.
In this case, one is led to a species scale of order~\cite{Montero:2022prj}
\begin{equation}
  \Lambda_{\rm sp} \sim M_p^{2/3} \ R_{\rm KK}^{-1/3} \sim M_p^{2/3} \ \Lambda^{1/12} \sim 10^9~{\rm GeV}
  \, .
\end{equation}

In this paper we explore the possibility
of a compact space with two dark dimensions of micron scale. In this case, the species scale
\begin{equation}
  \Lambda_{\rm sp} \sim (M_p / R_{\rm KK})^{1/2} \sim M_p^{1/2} \Lambda^{1/8}
\label{Lsp2}
\end{equation}
would be in the 10~TeV regime, within
reach of the Future Circular Collider in the modality of hadron-hadron interactions (FCC-hh)~\cite{FCC:2018vvp}. Moreover, this
scenario could simultaneously account for the cosmological and gauge
hierarchy problems. The goal of this article is to establish the required conditions for (\ref{Lsp2}) to
be consistent with experiment.

The layout of the paper is as follows. In Sec.~\ref{sec:2} we confront
(\ref{Lsp2}) to null results from collider searches, in
Sec.~\ref{sec:3} we confront it to 
astrophysical data, and in Sec.~\ref{sec:4} we confront it to cosmological
observations. In Sec.~\ref{sec:5} we reexamine whether neutrino masses and
mixings could exclusively occur due to physics in the bulk. In Sec.~\ref{sec:6} we show that an all-dark-matter
interpretation in terms of 6D primordial  black holes (PBH) should be
feasible for masses in the range  $10^8 \alt M_{\rm BH}/{\rm g} \alt
10^{21}$. In Sec.~\ref{sec:7} we take a look at the possibility of
probing $\Lambda_{\rm sp}$ with ultra-high-energy cosmic rays. Conclusions are given in Sec.~\ref{sec:8}.

\section{Collider Constraints}
\label{sec:2}

ATLAS and CMS are the two primary, general-purpose experiments, at the
Large Hadron Collider (LHC). Since the early days of the LHC the ATLAS
and CMS collaborations developed analysis strategies to search for
signals of large
extra dimensions.

A particular signal of these searches is the production
of a single very energetic ``mono-object'' that does not balance the
transverse momentum carried by anything else emerging from the
collision (as would be required by momentum and energy
conservation). Examples of such objects are very energetic photons or particle jets. Collisions
producing these mono-objects
only appear to be imbalanced because the emerging photon or jet is balanced by a graviton that escapes detection. As a
consequence, SM processes such as the production of a jet plus a $Z$ boson which decays
into neutrinos can mimic a graviton production signal. The absence of
any excess in the mono-photon or mono-jet event channels at ATLAS
and CMS has lead to stringent limits on the species scale, see
e.g.~\cite{CMS:2012lmn,ATLAS:2012ezx,CMS:2011esc,ATLAS:2011kno,CMS:2014jvv,ATLAS:2015qlt,ATLAS:2021kxv,CMS:2021far}. Null
results from searches of mono-objects were also reported by the Tevatron D0~\cite{D0:2008ayi} and
CDF~\cite{CDF:2008njt} collaborations.

One can also search for virtual graviton effects, which manifest
themselves as a new contribution to the continuum in the invariant
mass spectrum of two energetic photons or fermions (dileptons or
dijets). As of today no signals have been observed at ATLAS and CMS,
excluding such contributions for $\Lambda_{\rm sp}$ up to several TeV~\cite{CMS:2018ucw,ATLAS:2017ayi,CMS:2018dqv,CMS:2021ctt}.

Today, the most restrictive bounds on the species scale come from searches in events with an energetic jet and missing
transverse momentum. For $n=2$, the 95\% CL lower bound reported by the ATLAS Collaboration is~\cite{ATLAS:2021kxv}
\begin{equation}
  \Lambda_{\rm sp,min} = 4.5~{\rm TeV} 
\end{equation}
whereas the CMS Collaboration reported~\cite{CMS:2021far}
\begin{equation}
  \Lambda_{\rm sp,min} = 4.5~{\rm TeV} \, .
\end{equation}
We note that we have reduced the limits as given in in Table 10 of~\cite{ATLAS:2021kxv} and in Table 3
of~\cite{CMS:2021far} by a factor of $(2
\pi)^{n/(n+2)}$. This is because both the ATLAS~\cite{ATLAS:2015qlt} and CMS~\cite{CMS:2014jvv}
collaborations definine the volume of the compact space as introduced
in~\cite{Giudice:1998ck}, i.e., volume equal to $R_{\rm KK}^n$ rather
than $(2 \pi R_{\rm KK})^n$.

If the large extra dimension scenario 
 is embedded in a string theory at the TeV scale~\cite{Antoniadis:1998ig}, we expect the
 string scale
 $M_s \alt \Lambda_{\rm sp}$,  and thus production of string
 resonances at the LHC~\cite{Anchordoqui:2007da,Anchordoqui:2008ac,Lust:2008qc,Anchordoqui:2008di,Anchordoqui:2009mm,Anchordoqui:2014wha}. Only one assumption is necessary in order to
 set up a solid framework: the string coupling must be small in order
 to rely on perturbation theory in the computations of scattering
 amplitudes. In this case, black hole production and other strong
 gravity effects occur at energies above the string scale; therefore
 at least a few lowest Regge recurrences are available for
 examination, free from interference with some complex quantum
 gravitational phenomena.  Certain amplitudes to leading order in
 string coupling (but including all string $\alpha'$ corrections) are
 universal. These amplitudes, which include $2 \to 2$ scattering
 processes involving 4 gluons or 2 gluons and 2 quarks, are
 independent of the details of the compactification, such as the
 configuration of branes, the geometry of the extra dimensions, and
 whether SUSY is broken or not. This model independence makes it
 possible to compute the string corrections to dijet signals at the
 LHC~\cite{Anchordoqui:2008di}.

In addition, the mixing between hypercharge and the gauge baryon
number symmetry implies that tree level gluon-gluon scattering processes such as $gg \to g\gamma$ and $gg \to \gamma \gamma$ are allowed even though they can only appear at loop order in the SM~\cite{Anchordoqui:2007da}. Unlike the dijet case, these amplitudes are mildly model dependent, {\em i.e.,} they depend on a mixing parameter which is fixed by the D-brane model but are otherwise independent of the compactification scheme.

The ATLAS and CMS collaborations have searched for signals of Regge
recurrences analyzing the dijet and $\gamma$ + jet invariant mass
distributions~\cite{Khachatryan:2010jd,Chatrchyan:2011ns,ATLAS:2011ai,Chatrchyan:2013qha,Aad:2013cva,Khachatryan:2015sja,Khachatryan:2015dcf,Sirunyan:2016iap,Sirunyan:2018xlo,Sirunyan:2019vgj}. Null
results of these searches lead to a lower bound $M_s > 8~{\rm TeV}$
at the 95\% CL. Putting all this together, we conclude that collider
constraints are consistent with two dark dimensions of micron size and
$\Lambda_{\rm sp} \sim 10~{\rm TeV}$.

\section{Astrophysical Constraints} 
\label{sec:3}

 Supernova (SN) explosions have long been considered to be powerful
 probes of large extra dimensions~\cite{Arkani-Hamed:1998sfv}. This is because SN cores could emit
 sizable fluxes of KK gravitons, with masses up to about 100~MeV. The
 KK emission process could then compete with neutrino cooling, shortening the observable signal. This implies that the size of the extra
dimensions could be constrained by requiring that SN 1987A did not emit
more KK gravitons than is compatible with the observed neutrino signal
duration~\cite{Cullen:1999hc,Barger:1999jf,Hanhart:2000er,Hanhart:2001fx}. This
reasoning leads to a bound 
\begin{equation}
R_{\rm KK} < \left\{\begin{array}{ll}
490~{\rm m} &
              ~~~~~~{\rm for} \ n=1 \\
0.96~\mu{\rm m} & ~~~~~~{\rm for} \ n=2 \end{array} \right.
 \,, \label{bound}                        \end{equation}            
corresponding to $\Lambda_{\rm sp}> 740~{\rm TeV}$ and $\Lambda_{\rm sp} > 8.9~{\rm TeV}$, respectively~\cite{Hannestad:2003yd}.

A more restrictive constraint emerges by considering the radiative decay of KK
gravitons, which could produce
potentially observable gamma rays~\cite{Hannestad:2001jv}. The partial decay width of this
process is estimated to be
\begin{equation}
  \Gamma^{\vec l}_{\gamma \gamma} \sim {\tilde \lambda}^2 \ \frac{
    m^3_{\rm KK} \ \vec l^{\;3}}{80 \pi \
    M_p^2} \,,
\label{gammagamma}
\end{equation}
where we have considered a
tower of equally spaced gravitons, indexed by an integer $\vec l$ and
with mass  $m_{l} = \vec l \ m_{\rm KK}$, and where the parameter $\tilde \lambda$ measures the value of the dark graviton wave
function at the SM brane and is expected to be ${\cal
  O}(1)$~\cite{Han:1998sg,Hall:1999mk}.
Following~\cite{Hannestad:2003yd} we assume $\tilde
\lambda =1$. The non-observation of these gamma
  rays in data collected by the Energetic Gamma Ray Experiment
  Telescope (EGRET) leads to the following limits on the 
  compactification scale:
  \begin{itemize}[noitemsep,topsep=0pt]
    \item The diffuse $\gamma$-ray flux measured by the EGRET
      instrument constrains the number of KK-gravitons that may have
      been emitted by all cosmic SNe. This constraints can be
      translated into the bound
\begin{equation}
R_{\rm KK} < \left\{\begin{array}{ll}
4.9 \times 10^2~{\rm m} &
              ~~~~~~{\rm for} \ n=1 \\
1~\mu{\rm m} & ~~~~~~{\rm for} \ n=2 \end{array} \right.
 \,,                    \label{ocho}     \end{equation}            
corresponding to limits of $\Lambda_{\rm sp}> 3.4 \times 10^3~{\rm TeV}$ and $\Lambda_{\rm sp} > 28~{\rm TeV}$, respectively~\cite{Hannestad:2003yd}.
\item Cas A is a young SN remnant. Based on its age, a cloud of
  emitted KK-gravitons should still appear as a point source to
  EGRET. However, EGRET does not observe a photon flux at the expected
  point in space. The difference between the theoretically expected flux and observation restricts the emissivity of KK gravitons. This
  restriction can be translated into the bound
  \begin{equation}
R_{\rm KK} < \left\{\begin{array}{ll}
5.3 \times 10^3~{\rm m} &
              ~~~~~~{\rm for} \ n=1 \\
1.3~\mu{\rm m} & ~~~~~~{\rm for} \ n=2 \end{array}
\right. \label{nueve}
 \,,                         \end{equation}            
corresponding to limits of $\Lambda_{\rm sp}> 330~{\rm TeV}$ and
$\Lambda_{\rm sp} > 7.7~{\rm TeV}$,
respectively~\cite{Hannestad:2003yd}.
\item In a core collapse SN most KK gravitons are produced close to the
kinematic threshold. For SN core temperature $\sim 30~{\rm
  MeV}$, the threshold condition implies that the typical mass of KK
gravitons is ${\cal O}(100~{\rm MeV})$. Thus, most gravitons leave the SN core with rather
non-relativistic velocities such that a large fraction of them ends up
being gravitationally retained in a cloud around the neutron star
remnant. Trapped in this cloud the KK gravitons would subsequently
decay to SM particles on a time scale comparable to the age of the
universe. We would then expect that neutron stars should shine
brightly in 100~MeV gamma-rays. However, this is not the case relative to  EGRET data, yielding 
\begin{equation}
R_{\rm KK} < \left\{\begin{array}{ll}
16~{\rm m} &
              ~~~~~~{\rm for} \ n=1 \\
67~{\rm nm} & ~~~~~~{\rm for} \ n=2 \end{array} \right.
 \,,                 \label{diez}        \end{equation}            
corresponding to limits of $\Lambda_{\rm sp}> 2.2 \times 10^3~{\rm TeV}$ and
$\Lambda_{\rm sp} > 34~{\rm TeV}$,
respectively~\cite{Hannestad:2003yd}.
\end{itemize}
The decay of KK gravitons  also leads to an excess heating of neutron
stars~\cite{Hannestad:2001xi}. This heating effect should have been seen by the Hubble Space
Telescope which is able to observe the thermal emission from the
surface of neutron stars. The lack of such an excess heat leads to the
most stringent bounds on the size of the compact space,
\begin{equation}
R_{\rm KK} < \left\{\begin{array}{ll}
8.3~{\rm m} &
              ~~~~~~{\rm for} \ n=1 \\
59~{\rm nm} & ~~~~~~{\rm for} \ n=2 \end{array} \right.
 \,,                 \label{once}        \end{equation}  
corresponding to $\Lambda_{\rm sp}> 2.8 \times 10^3~{\rm TeV}$ and
$\Lambda_{\rm sp} > 36~{\rm TeV}$, respectively~\cite{Hannestad:2003yd}.

Notwithstanding, the constraints
coming from the decays of KK gravitons into photons are model
dependent and can be completely evaded. Note that the constraints from EGRET data and neutron-star heating rely on the assumption that KK modes can
only decay into SM fields localized on the brane, but
cannot decay into other KK modes with smaller bulk momenta. If large
extra dimensions do not admit isometries this assumption is in general
not valid and the bounds in (\ref{ocho}), (\ref{nueve}), (\ref{diez}),
and (\ref{once}) can be evaded~\cite{Mohapatra:2003ah}. Indeed, if the KK momentum of the graviton tower is not conserved a given KK mode of the tower could decay into final states that include other,
lighter KK excitations. 

Following~\cite{Gonzalo:2022jac}, we set $n=1$ and assume that a parent particle with mass $m_l$ can decay to
two daughter particles with masses $m_{l'}$ and $m_{l''}$ such that
$m_l = m_{l'} + m_{l''} + \epsilon$, with $\epsilon \leq 
m_{\rm KK}\delta \ $ and $\delta \sim {\cal O} (1)$. As
might be expected, the KK mode $l$ decays with gravitational
strength to lighter elements of the tower and so the partial decay
width is shown to be
\begin{equation}
\Gamma^{l}_{l' l''}
\sim m_l^3/M_p^2 \, .
\end{equation}
In principle, there are $m_l/m_{\rm KK}$ possibilities for
the  $l' l''$  pair of KK modes that $l$ can decay to, but because of
a small violation of KK quantum number,
 the choice effectively becomes
$m_l \delta/m_{\rm KK}$. Taking into account the phase space factor which
is roughly the velocity of decay products at threshold,
\begin{equation}
  v \sim \sqrt{m_{\rm KK} \ \delta /m_l}\,,
\label{velocity}
\end{equation}
  the total
  decay width of graviton $l$ is found to be, 
\begin{eqnarray}
  \Gamma^{l}_{\rm tot} & \sim & \sum_{l'<l} \ \ \sum_{0<l''<l-l'}
                              \Gamma^{l}_{{l'} {l''}} \nonumber \\
  & \sim &
  \beta^2 \frac{m_l^3}{M_p^2} \times \frac{m_l}{m_{\rm KK}} \ \delta
   \times \sqrt{\frac{m_{\rm KK} \ \delta }{m_l} } \nonumber \\
  & \sim & \beta^2 \
    \delta^{3/2} \frac{m_l^{7/2}}{M_p^2 m_{\rm KK}^{1/2}} \,,
\label{Gtot}
\end{eqnarray}   
 where $\beta$ is parameter that controls the strength of the intra-tower decay amplitudes which correlates with the amplitudes on inhomogeneities in the dark dimension~\cite{Gonzalo:2022jac}. Assuming that for times larger than $1/\Gamma^{_l}_{\rm tot}$ dark matter
which is heavier than the corresponding $m_l$ has already decayed, it
is straightforward to see that
\begin{equation}
  m_l (t) \sim \left(\frac{M_p^4 \ m_{\rm KK}}{\beta^4 \ \delta^3}\right)^{1/7} t^{-2/7} \,,
\label{mevol}
\end{equation}
where $t$ indicates the time elapsed since the big bang.

For $n=2$,
conservation of energy forces the
the momenta of the decay products to be almost parallel and so the
number of decay channels available is still effectively 1D, up to a
width that will be set by $\delta$. The  evolution of the
KK mass in the graviton tower is also given by (\ref{mevol}).

We now turn to discuss how dark-to-dark decays provide a path to
evade the lower bounds
on the species scale given in (\ref{ocho}), (\ref{diez}),
and (\ref{once}). Following Hannestad and Raffelt (HR), we parametrize the
bound on the species scale in the absence of intra-tower decays by
\begin{equation}
\Lambda_{\rm sp,min}^{\rm HR} = \left(\frac{f_{\rm KK}^{\rm
      HR}}{0.5}\right)^{-1/(n+2)} \ \Lambda_{{\rm sp},n}^{\rm SN\,
  1987A} \ ,
\label{fHR}
\end{equation}
where $f_{\rm KK}^{\rm HR}$ is the
fraction of the total energy emitted by the SN in the form of KK
gravitons and $\Lambda_{{\rm sp},n}^{\rm SN\, 1987A}$ is the bound on
the species scale derived from observations of SN 1987A~\cite{Hannestad:2001xi}.  In the absence of dark-to-dark decays all of the $f_{\rm
  KK}^{\rm HR}$ fraction would decay dominantly into photons, and so
the null results on searches for these photons further reduce the
fraction of allowed emitted KK modes, i.e.
$f_{\rm KK}^{\rm HR} < 0.5$.
For example, for $n=2$, the bound in (\ref{fHR}) can be recast as
\begin{equation}
\Lambda_{\rm sp,min}^{\rm HR} = \left(\frac{f_{\rm KK}^{\rm
      HR}}{0.5}\right)^{-1/4} 8.9~{\rm TeV} \, .
\end{equation}

Due to
the intra-tower decays the fraction of KK gravitons that could decay
into photons gets a multiplicative correction to accomodate the
evolution of dark-to-dark decays,
\begin{equation}
         f_{\rm KK}^{\rm new} = f_{\rm KK}^{\rm HR} \times \overline{f_{\rm KK}}
\end{equation}
where $\overline{f_{\rm KK}}$ is determined by the ratio of the
initial decay rate $\Gamma_0$ of the KK gravitons to photons to the
final decay rate $\Gamma (t)$ of the KK gravitons to photons, 
\begin{equation}
          \overline{f_{\rm KK}} = \int_0^{\rm \tau_{\rm NS}}
          \frac{\Gamma_0}{\Gamma (t)} \ 
          \frac{dt}{\tau_{\rm NS}} \,,
\label{int}
        \end{equation}
     and where $\tau_{\rm NS} \sim 17~{\rm Myr}$ is the lifetime of the neutron star~\cite{Hannestad:2003yd},
\begin{equation}        
 \Gamma_0 \sim m_0^3/M_p^2 \,,
\label{Gamma0}
\end{equation}
$m_0 \sim 100~{\rm MeV}$ is the mass of the KK modes at
production, and 
\begin{equation}
  \Gamma(t) = m_l^3(t)/M_p^2  \, .
\label{Gammat}
\end{equation}  
Substituting (\ref{Gamma0}) and (\ref{Gammat}) into (\ref{int}) while
using (\ref{mevol}) we obtain
\begin{equation}
\overline {f_{\rm KK}} \sim \frac{13}{7} \left[\frac{m_0}{(M_p^4 \ m_{\rm
    KK})^{1/7}} \right]^3 \tau_{\rm NS}^{6/7} \sim 881 \,,
\end{equation}
where we have taken $\delta = \beta =1$. The corrected lower limit on the species scale is then 
\begin{equation}
\Lambda_{\rm sp,min}^{\rm new} =  \left(\frac{f_{\rm KK}^{\rm HR}}{0.5}\right)^{-\frac{1}{n+2}} \times  \overline {f_{\rm
  KK}}^{-\frac{1}{n+2}} \  \Lambda_{{\rm sp},n}^{\rm SN\, 1987A}  \, .
\end{equation}  
All in all, for $n=2$, the new bounds on the species scale associated to (\ref{diez})
and (\ref{once}) become 
\begin{equation}
\Lambda_{\rm sp,min}^{\rm new} = \overline{f_{\rm KK}}^{-1/4} \ 34~{\rm TeV} =
6.2~{\rm TeV}
\end{equation}
and 
\begin{equation}
\Lambda_{\rm sp,min}^{\rm new} = \overline{f_{\rm KK}}^{-1/4} \
36~{\rm TeV} =
6.6~{\rm TeV} ,
\end{equation}
respectively.

Duplicating this procedure, but substituting $\tau_{\rm NS}$ by the
age of the universe in the limit of integration of (\ref{int}), we obtain the
$\overline{f_{\rm KK}}$ to rescale the bound on the species scale
associated to (\ref{ocho}). It is straightforward to see that the
corresponding value of $\overline{f_{\rm KK}} > 881$ and therefore the
bound on (\ref{ocho}) can be evaded.

We conclude that in the absence of isometries the SN
 constraints are given by (\ref{bound}) and one or two extra 
 dimensions with $R_{\rm KK} \sim 1~\mu{\rm m}$ are consistent with observations.

\section{Cosmological Constraints}
\label{sec:4}

Cosmological observations lead to additional model independent and model
dependent constraints. We begin discussing constraints on models in
which KK intra-tower decay is not allowed, and then we relax
this condition by considering extra dimensions which do not admit isometries. 

Model independent constraints originate again in the emission of KK
modes. Expansion dominated cooling gives a model independent
bound. Indeed, an upper limit on the ``normalcy'' temperature $T_*$ at which the
universe must be free of bulk modes can be derived by demanding that
the rate at which radiation energy density on the brane evaporates
into KK modes in the bulk remains below the normal cooling due to cosmological expansion,
\begin{equation}
  T_* \alt 10^{\frac{7n-7}{n+1}} \ \left(\frac{\Lambda_{\rm
        sp}}{10~{\rm TeV}} \right)^{\frac{n+2}{n+1}}~{\rm MeV}\,,
\end{equation}
which for $n=2$, leads to $T_* \alt 200~{\rm  MeV}$~\cite{Arkani-Hamed:1998sfv}. 
Big Bang nucleosynthesis (BBN) set a more restrictive (but model
dependent) bound on $T_*$. From the 4D perspective, the gravitons
produced at temperature $T$ are massive KK modes with mass $\sim
T$. This is because the emission rate of each KK mode $\sim
T^3/M_p$. It goes without saying that for given $T$, all KK
modes up to $T$ are produced, but most particles have a mass $\sim T$,
because the emission rate goes down rapidly with decreasing
$T$. Putting all this together, the energy density in KK modes redshifts away as $a^3$ rather than
$a^4$, where $a$ is the cosmic scale factor. Then, to insure normal
expansion rate during BBN the bound on the normalcy temperature is
slightly stronger,
\begin{equation}
  T_* \alt 10^{\frac{7n-7}{n+2}} \ \left(\frac{\Lambda_{\rm
        sp}}{10~{\rm TeV}} \right)~{\rm MeV}\,,
  \label{BBN-bound}
\end{equation}
which for $n=2$, leads to $T_* \alt 50~{\rm MeV}$~\cite{Arkani-Hamed:1998sfv}.

Even more restrictive bounds emerge by requiring that the energy
density of KK gravitons today does not overclose the universe. Indeed,
to overcome overproduction of KK modes at early times the temperature
at which the universe enters the radiation dominated epoch must be
delicately fine-tuned~\cite{Arkani-Hamed:1998sfv,Hall:1999mk,Hannestad:2001nq}. The contribution of thermally produced KK to the energy density today
is estimated to be
\begin{equation}
\frac{\rho_{{\rm KK}_0}}{T_0^3} \sim \frac{2}{3} \frac{M_p T_*^4}{\sqrt{g_*} \Lambda_{\rm
    sp}^4} \,,
\end{equation}
where $g_*$ is the effective number of relativistic degrees of freedom  and where we have used the fact that the ratio $\rho_{\rm KK}/T^3$ is
invariant. The critical density of the universe today corresponds to
$(\rho_{\rm crit}/T_0^3) \sim 3 \times 10^{-9}~{\rm
  GeV}$. Setting $\rho_{\rm KK} \leq \rho_{\rm crit}$, with
$\Lambda_{\rm sp} = 10~{\rm TeV}$ and $g_* =
10.75$ leads $T_* \sim
2.8~{\rm MeV}$~\cite{Arkani-Hamed:1998sfv}. This normalcy temperature
derived more carefully is closer to the decoupling temperature for the
muon and tau neutrinos. Indeed, for $T_* = 2.15~{\rm MeV}$, it follows
that
\begin{equation}
  R_{\rm KK} < 3.3 \ h~\mu{\rm m}  \sim 2.3~\mu{\rm m} \,,
\label{Rcosmobound}
\end{equation}
which corresponds to
\begin{equation}
  \Lambda_{\rm sp} > 5~{\rm TeV}/\sqrt{h}  \sim 6~{\rm TeV}\,,
  \label{Lcosmobound}
\end{equation}  
where $H_0 = 100\, h~{\rm km \, Mpc^{-1} \, s^{-1}}$ is the Hubble
constant, with $h \simeq 0.7$~\cite{Hall:1999mk}.\footnote{Note that $M = 2^{1/(n+2)} (2 \pi)^{n/(n+2)} \Lambda_{\rm
    sp}$~\cite{Hannestad:2003yd}, where $M$ is the underlying $(n +
  4)$-dimensional scale adopted in~\cite{Hall:1999mk}.} Needless to
say, the bounds in (\ref{Rcosmobound}) and (\ref{Lcosmobound}) only take into account thermal production of the
KK modes of gravitons in the early universe. Successful inflation and
reheating, as well as baryogenesis, usually stand in need for the
existence of fields in the bulk, most notably the inflaton. The
non-thermal production of KK modes accompanying the inflaton decay
would further constrain $R_{\rm KK}$ and exclude the possibility of two
extra dimensions of micron size~\cite{Hannestad:2001nq}.\footnote{For
  values of $T_*$ above the QCD phase transition further
  considerations are necessary to constrain $R_{\rm KK}$~\cite{Fairbairn:2001ct,Macesanu:2004gf}.}

If we are prepared to accept that KK gravitons are only produced
thermally, it is of interest to further investigate whether low
reheating temperature universes, with $T_* \sim 2~{\rm MeV}$, remain
consistent with cosmological observations. Lower bounds on the reheating
temperature have been derived assuming the the late-time entropy
production near BBN is dominated by radiative decays. The incomplete
thermalization of neutrinos could modify the production of $^{4}$He
and this leads to a bound $T_* > 0.5~{\rm MeV}$ at
95\%CL~\cite{Kawasaki:1999na}. When
hadronic decays are also considered $T_* > 2.4~{\rm MeV}$ at
95\%CL~\cite{Kawasaki:2000en}. In addition, the effect of flavor
neutrino oscillations turns out to be quite relevant to accommodate BBN, shifting the
lower bound to $T_*$ higher temperatures.  Indeed, the thermalization process
of two- and three-flavor neutrino oscillations have been analyzed yielding
the following 95\% CL bounds:
$T_* > 2~{\rm MeV}$ and $T_* > 4.1~{\rm MeV}$ assuming radiative decay
of the massive
particles~\cite{Ichikawa:2005vw,deSalas:2015glj}. More recently, it
was shown that when both neutrino oscillations and
self-interactions are considered in the thermalization process the
95\% CL lower bounds on the normalcy temperature are less restrictive: $T_*>
1.8~{\rm MeV}$ and $T_*> 4~{\rm MeV}$, for radiative decays and
hadronic decays; respectively~\cite{Hasegawa:2019jsa}. CMB data and
large scale structure lead to comparable bounds on the normalcy
temperature, $T_* \agt 2~{\rm
  MeV}$~\cite{Hannestad:2004px,Ichikawa:2006vm,DeBernardis:2008zz}.\footnote{We
  note in passing that a recent study combining Planck CMB data and
  DESI DR1 results with  $N_{\rm eff} = 2.58$  gives $T_* > 3.79~{\rm MeV}$ at
  95\% CL, and when BBN predictions are included in the likelihood
  analysis using $N_{\rm eff} = 2.98$ yields $T_* > 5.96~{\rm
    MeV}$~\cite{Barbieri:2025moq}. However, it should also be noted
  that because both the primordial helium abundance $Y_{\rm P}$ and
  the relativistic degrees of freedom affect the CMB damping tail,
  they are partially degenerate. Allowing $N_{\rm eff}$ and $Y_{\rm
    P}$ to vary in the likelihood analysis leads to weaker lower bound of
  $N_{\rm eff} = 2.32$~\cite{Planck:2018vyg}. This will naturally
  shift the value of the minimum allowed $T_*$ to lower temperatures.}

\begin{figure}[tbh]
  \postscript{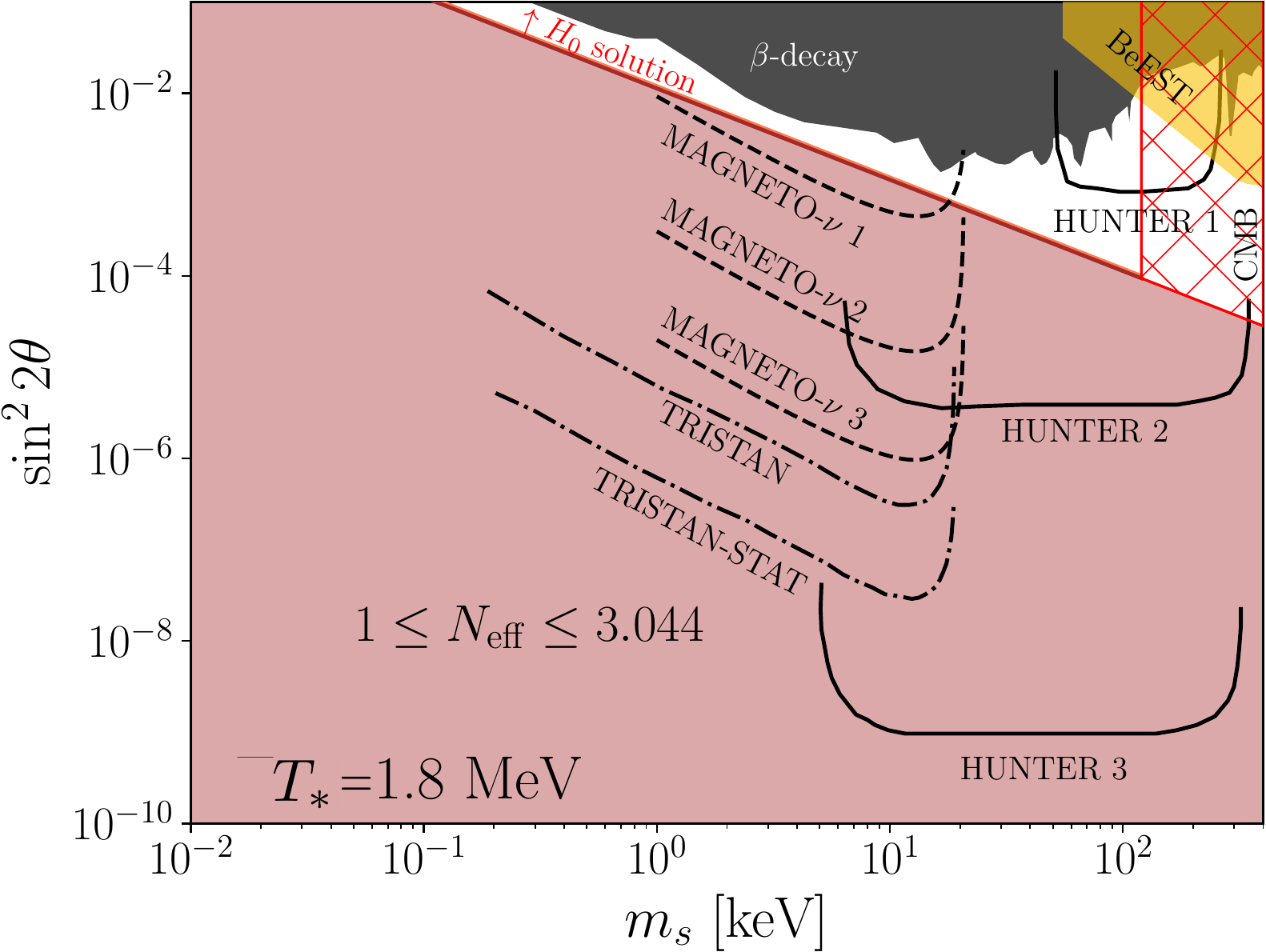}{0.95}
  \caption{The  $(m_s, \theta)$ parameter space for
    $T_* = 1.8~{\rm MeV}$. The black region shows constraints from a
    collection of nuclear beta decay searches~\cite{ParticleDataGroup:2016lqr} and the region in golden
    yellow constraints from $\beta$-decay searches at
    BeEST~\cite{Friedrich:2020nze}. The thermal nature of the
CMB constrains the red hatched portion in the upper
right~\cite{Hu:1993gc}. In the pink region $1 \leq N_{\rm eff} \leq 3.044$. The diagonal red line correspond to the scenario when
the decay happens at the temperature of matter-radiation
equality. Above the diagonal red line, $N_{\rm eff} > 3.044$ and the parameter space can accommodate an alleviation
of the $H_0$ tension. The experimental sensitivity of HUNTER~\cite{Smith:2016vku},
TRISTAN~\cite{KATRIN:2018oow}, MAGNETO-$\nu$~\cite{Lee:2025txi}, and PTOLEMY~\cite{PTOLEMY:2019hkd} for nuclear decay
searches of keV-scale sterile neutrinos
is also
shown. Taken from~\cite{Abazajian:2023reo}.
\label{fig:1}}
\end{figure}

Recently, cosmologies with ${\cal O} ({\rm MeV})$ normalcy
temperatures have become the focus of a great deal of interest because
they can accommodate several new physics scenarios that would normally
be constrained by high-temperature reheating models, including massive
sterile neutrinos~\cite{Abazajian:2023reo}. For example, if $T_* \sim
1.8~{\rm MeV}$ sterile
neutrinos (of mass $m_s \sim {\rm keV}$) with large mixing angle $\theta$
and a dark decay to radiation may ameliorate the $H_0$ tension~\cite{Escudero:2022rbq}. Several neutrino experiments are
scanning the 0.1 to 100~keV mass range to search for these sterile
neutrinos, which may be visible in the lab but may be invisible to
cosmological observations; for details see Fig.~\ref{fig:1}.

For $n$ extra dimensions, the constraint from cosmological
overproduction of bulk graviton modes can be written as
\begin{equation}
  T_* \sim 10^{3 + (6n-15)/(n+2)}   \left(\frac{\Lambda_{\rm
        sp}}{10^9~{\rm GeV}} \right)~{\rm GeV} \,,
\end{equation}
which implies that for a single dark dimension in the micron range the universe must be free of bulk modes
at $T_* \sim 1~{\rm GeV}$~\cite{Arkani-Hamed:1998sfv}.

Dark gravitons (KK modes) provide a promising dark matter
candidate~\cite{Gonzalo:2022jac}. As noted in the previous section, the cosmological evolution of the dark sector is mostly driven by
dark-to-dark decay processes, which regulate the decay of KK gravitons
within the dark tower,
realizing a particular version of the dynamical dark matter
framework~\cite{Dienes:2011ja}. CMB and Lyman $\alpha$ data set constraints
on $\tilde \lambda$~\cite{Law-Smith:2023czn}. In addition, KK decays impart momentum to
the daughter particles, injecting kinetic energy into virialized
self-gravitating dark-matter halos. These momentum
``kicks'' -- which have been shown in (\ref{velocity}) to depend on $\delta$ ---
gradually reduce the central density of
cuspy dark matter halos and deplete the amount of mass in the halos~\cite{Peter:2010jy,DES:2022doi}. This process also makes
sub-halos that have a virial velocity smaller than the velocity kick  more
susceptible to tidal disruption as they orbit within a host halo,
leading to the suppression of the abundance of low-mass halos and
sub-halos at late times. To preserve the structure and stability of
galaxy halos, the analysis of~\cite{Obied:2023clp} severely constrains the
$(\tilde \lambda,\delta,\beta)$ parameter space. Using Fig.~4 in~\cite{Obied:2023clp} it is
easily seen that the allowed region of the $(\tilde
\lambda,\delta,\beta)$ parameter space leads to larger values of $\overline{f_{\rm KK}}$.

In summary, two extra dimensions of micron-size require an upper limit on
the normalcy temperature at which the universe must be free of bulk
modes, which is estimated to be $T_* \sim 2.15~{\rm MeV}$~\cite{Hall:1999mk}. This is
comparable to the lowest possible
reheating temperature consistent with CMB data
and abundances predicted by BBN, which is estimated to be $T_* \sim 1.8~{\rm MeV}$~\cite{Hasegawa:2019jsa}.

\section{Constraints on Neutrino Towers}
\label{sec:5}

The smallness of the neutrino mass may be ascribed to the fact that
right-handed ($R$) neutrinos could live in the
bulk~\cite{Dienes:1998sb,Arkani-Hamed:1998wuz,Dvali:1999cn}. The
coupling of $R$-neutrinos to the left-handed SM neutrinos living on the brane is
inversely proportional to the square-root of the bulk volume. The
Dirac neutrino mass can be shown to be
\begin{equation}
  m_\nu \sim y \ \langle H \rangle \ \Lambda_{\rm sp}/M_p \,,
\end{equation}  
where $y$ is the Yukawa coupling and $\langle H \rangle = 246~{\rm
  GeV}$ is the Higgs vacuum expectation
value. Note that for
$\Lambda_{\rm sp}/M_p \ll 1$, the neutrino mass scale is well below
the electroweak symmetry breaking scale $\langle H \rangle$, even for
order-one Yukawa couplings.

If $R$-neutrinos live in the bulk their KK modes would mix with SM
neutrinos modifying their oscillation pattern. The phenomenology is
similar to that of light sterile neutrinos, which induce observable
short-baseline neutrino oscillations and small perturbations to the
neutrino oscillations in solar, atmospheric, and long-baseline
experiments that are well-described by standard three-neutrino
mixing. The non-observation of these effects in neutrino-detection facilities set  90\%~C.L. bounds on the size
of the largest extra dimension
\begin{equation}
R_{\rm KK} < \left\{\begin{array}{ll}
 0.2~\mu{\rm m} &
              ~~~~~~{\rm NO}  \\
0.1~\mu{\rm m} & ~~~~~~{\rm IO} \end{array} \right.
 \,,                 \label{nubound}        \end{equation}  
for normal (NO) and inverted (IO) ordering of neutrino
masses~\cite{Davoudiasl:2002fq,Machado:2011jt,Forero:2022skg}. However, these bounds could be significantly relaxed in the
presence of bulk neutrino masses, because the mixing of the first KK
modes to active SM neutrinos becomes suppressed, making the
contribution of heavier KK modes to oscillations relatively more important~\cite{Carena:2017qhd,Anchordoqui:2023wkm}.

If $R$-neutrinos live in the bulk, the limit (\ref{bound}) would be stronger
because neutrino towers could offer more modes into which energy may
be lost. The SN energy loss rate because of nucleon-nucleon
gravi-strahlung $\dot \varepsilon_{\rm SN} \propto \Lambda_{\rm
  sp}^{-(n+2)}$~\cite{Cullen:1999hc}. A rough estimate of the
$\Lambda_{\rm sp}$ lower bound can be established by scaling the graviton energy loss
rate of the total number of degrees of freedom,
$N$, in the bulk whose couplings allow them to be
emitted singly starting only from brane states
\begin{equation}
  \Lambda_{\rm sp} \agt 8.9 \ \left(\frac{N}{3}\right)^{1/4}~{\rm TeV} \,,
\end{equation}
with $n=2$~\cite{Burgess:2004yq}. Obviously, if only
gravitons could be emitted, then we have $N = 3$ because out of the 9
6D graviton polarizations only 3 can couple to purely 4D stress
energy. Each Weyl $R$-neutrino contributes with two polarizations,
yielding $\Lambda_{\rm sp} \agt 12~{\rm TeV}$. For $n=1$, the cosmological evolution of $R$-neutrino towers was discussed
elsewhere~\cite{Anchordoqui:2024xvl}.

\section{Six-Dimensional Black Holes\\ as All Dark Matter}
\label{sec:6}

It is well-known that black holes are expected to have formed when
very large density perturbations collapse~\cite{Zeldovich:1967lct,Hawking:1971ei,Carr:1974nx}. An attractive idea is that
these overdensities are of inflationary origin~\cite{Hodges:1990bf,Carr:1993aq, Ivanov:1994pa}. Once the overdense
regions come back into causal contact after inflation, they instantly
collapse if they exceed a medium-specific threshold. These PBHs fulfill all of the necessary
requirements to be a captivating dark matter candidate: they are cold,
non-baryonic, quasi-stable, and can be formed in the right abundance
to be the dark matter. Actually, if PBHs contribute to more than 10\%
of the dark matter density, then their energy density today is of the
same order as that of the baryons~\cite{Wu:2021gtd}. This cosmic coincidence might hint
at a mutual origin for PBHs and the baryon
asymmetry of the Universe. Strikingly, PBHs fostered by a transition from
a slow-roll to ultraslow-roll during single field inflation can
contribute as a significant dark matter component~\cite{Leach:2001zf}, while such a transition of the inflationary background can trigger successful baryogenesis via the Afleck-Dine mechanism~\cite{Affleck:1984fy}. 

PBHs evaporate by emitting Hawking radiation, with a temperature inversely
proportional to its mass~\cite{Hawking:1974rv,Hawking:1975vcx}. Nevertheless,
a PBH could be stable on cosmological scales and its lifetime could
even be longer than the age of
the Universe. For example, if the initial mass of a 4D PBH is
$M_{\rm BH} \agt 10^{15}~{\rm g}$ it can survive until today~\cite{Page:1976df}. 
Even so, an all-dark-matter interpretation in terms of PBHs have been
ruled out through observations across most mass ranges~\cite{Carr:2020xqk,Green:2020jor,Villanueva-Domingo:2021spv}. Constraints mostly come from microlensing
surveys and non-observation of Hawking radiation. PBHs in the mass
range $10^{17.5} \alt M_{\rm BH}/{\rm g} \alt 10^{21}$ remain a viable explanation for all dark matter. 

Higher-dimensional black holes of Schwarzschild radii smaller than a micron are: bigger,
colder, and longer-lived than a usual 4D black hole of the same mass~\cite{Argyres:1998qn}. In a series of recent publications we have used these black hole properties
to extend the PBH range as the entirety of dark matter compared to that in the 4D theory by
several orders of magnitude in the low mass window~\cite{Anchordoqui:2022txe,Anchordoqui:2022tgp,Anchordoqui:2024akj,Anchordoqui:2024dxu}. In what follows we
particularize  our investigation to 6D black holes.

Shortly after being formed a higher-dimensional PBH would emit SM particles and gravitons on the brane as well as gravitons into the
bulk. However, the recoil effect due to graviton emission imparts the black hole a relative kick
velocity with respect to the brane, allowing the PBH to wander off
into the
bulk~\cite{Frolov:2002as,Frolov:2002gf,Flachi:2005hi,Flachi:2006hw,Anchordoqui:2024jkn}. The Hawking evaporation time characterizing the lifetime
of such a bulk PBH is estimated to be
\begin{equation}
  \tau_{\rm BH} \sim r_s \ S_{\rm BH}  \sim \frac{1}{\Lambda_{\rm sp}} \ \left(\frac{M_{\rm
      BH}}{\Lambda_{\rm sp}}\right)^{(n+3)/(n+1)} \, ,
\label{lifetime}
\end{equation}
where
\begin{equation}
r_s \sim \frac{1}{\Lambda_{\rm sp}} \ \left(\frac{M_{\rm
      BH}}{\Lambda_{\rm sp}}\right)^{1/(n+1)}
\end{equation}
  is
  the $(4+n)$-dimensional Schwarzschild radius~\cite{Tangherlini:1963bw,Myers:1986un} and
\begin{equation}
  S_{\rm BH} \sim (M_{\rm BH}/\Lambda_{\rm sp})^{(n+2)/(n+1)}
\end{equation}
is the black hole entropy~\cite{Anchordoqui:2001cg}. All in all, taking $n=2$ and
$\Lambda_{\rm sp} \sim 10~{\rm TeV}$ we obtain
\begin{equation}
  \tau_{\rm BH} \sim 13.7 \left(\frac{M_{\rm BH}}{10^{7.8}~{\rm
        g}} \right)^{5/3}~{\rm Gyr} \, .
\end{equation}  
We conclude that the mass range for 6D PBHs to make all the
cosmological dark matter is then 
\begin{equation}
  10^{8} \alt M_{\rm BH}/{\rm g} \alt 10^{21} \, .
\end{equation}
However, one caveat of the PBH all dark-matter interpretation is that for $R_{\rm KK} \sim 1~\mu{\rm m}$, thermal production
of KK gravitons almost overcloses the universe. Therefore, a significant fraction of dark matter
in the form of PBHs would exacerbate the required fine tuning on the reheating temperature.

In closing, we stress that if there were 6D primordial near-extremal
black holes in nature, then it would be possible to lower the minimum
mass allowing a PBH all-dark-matter interpretation, because
near-extremal black holes are colder and longer-lived~\cite{Anchordoqui:2024akj}. Note,
however, that near-extremal black holes cannot escape from the brane
into the bulk if they carry e.g., electromagnetic charge.

\section{Ultra-High-Energy Cosmic Rays\\ ~~~~~~~as Probes of Two Dark Dimensions}
\label{sec:7}

Extensive air showers induced from ultra-high-energy cosmic rays (UHECRs)
provide a window into understanding the most energetic interactions~\cite{Anchordoqui:2018qom}. To be specific, the
highest energy cosmic rays interact in the Earth’s atmosphere through
collisions with center-of-mass energies  $\sqrt{s} \sim 300~{\rm TeV}$, {\it viz.} not only well above 
possibilities at the Large Hadron Collider, but even beyond the projections for the
FCC-hh.

The initial interaction of a cosmic ray  in the atmosphere produces a set of secondary particles (mostly $\pi$'s) carrying a fraction of the
primary energy. The $\pi^0$'s produced in the first
interaction ($\sim 1/3$ of all pions in accord with isospin
invariance) promptly decay
to a pair of gamma rays. The gamma rays produce $e^\pm$ pairs when passing near nuclei. The electrons and positrons re-generate gamma rays via bremsstrahlung, thereby building an electromagnetic cascade. The $\pi^\pm$,
which carry $2/3$ of the energy lost by the cosmic ray, begin to move
through the atmosphere and either decay or interact generating new
sets of secondaries. In these interactions the $\pi^\pm$ again carry $2/3$ of the parent energy and $1/3$ goes into
electromagnetic cascade. After $j$ hadronic cascade generations, only
$(2/3)^j$ of the total energy remains in the hadronic cascade. The $\pi^{\pm}$ produce air shower muons when they
decay. The number of shower muons depends on the amount of energy that
is left in the hadronic cascade when pion energies have dropped to the
level where decay is more likely than collision. If this happens after
relatively few cascade generations, then copious muon production
occurs. If the reduction of pion energies takes relatively many generations, then more of the energy will have been lost from the
hadronic cascade to the electromagnetic cascade, and meager muon
production occurs. Eventually, the energy of the shower particles is
degraded to the point where ionization losses dominate, and their
number starts to decline. The electromagnetic cascade dissipates about 90\% of the
primary cosmic ray energy and the remaining 10\% is carried by muons
and neutrinos. The total number of muons produced by a cosmic ray
nucleus scales roughly as
\begin{equation}
  N_\mu \propto A \ (E/A)^{0.93} \,,
\label{Nmu}
\end{equation}
where $A$ is the nucleus baryon number~\cite{Anchordoqui:2018qom}. The expectation is then that a cosmic ray nucleus to
produce about $A^{0.07}$ more muons than a proton. This means that a shower initiated by an iron nucleus produces about 30\% more muons than a proton shower.

Measurements from several cosmic-ray experiments seem to indicate
there is a
significant, yet unexplained, discrepancy between the observed muon content in cosmic ray showers and that predicted by state-of-the-art interaction models, suggesting a need for refinements in our understanding of fundamental physics. In particular, Auger data suggest that showers induced by cosmic rays of energy $10^{9.8} < E/{\rm GeV}
< 10^{10.2}$ contain about 30\% to 60\% more muons than
expected~\cite{PierreAuger:2014ucz,PierreAuger:2016nfk}. The
significance of the discrepancy between data and model prediction has
been experimentally established at $8\sigma$~\cite{Albrecht:2021cxw}. The onset of the discrepancy is seen in showers with $E \sim
10^{7.7}~{\rm GeV}$, which corresponds to a first interaction at 
$\sqrt{s} \sim 10~{\rm TeV}$ in the nucleon-nucleon
system.

A thought-provoking observation is that scattering processes above the
species scale necessarily have internal structure in the compact space and to a
certain extent would
lead to the production of KK
modes. Indeed the phase space seen
by the emitted graviton grows quite quickly with energy on very
general grounds, yielding prolific production of KK modes in particle
interactions.\footnote{Moreover, black holes are expected to be produced in collisions
with $\sqrt{s} \gg \Lambda_{\rm sp}$~\cite{Banks:1999gd}. However, one caveat is that  QCD
cross sections dominate over the black hole production cross section
by a factor of roughly $10^9$~\cite{Anchordoqui:2003jr}. This implies  
that black holes produced in cosmic ray air showers seem effectively
unobservable. This is also the case for the production of string
massive modes~\cite{Cornet:2001gy}.} Along this line, air shower
simulations seem to indicate 
that production of KK gravitons by nucleon-nucleon bremsstrahlung (at
large impact parameter) 
could lead to a sizable fraction of shower missing energy~\cite{Kazanas:2001ep,Koch:2006yia}.

\begin{figure}[tbh]
  \postscript{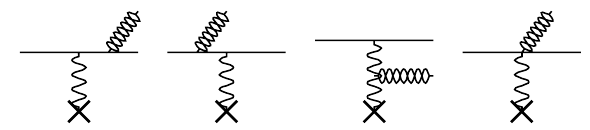}{0.95}
  \caption{Feynman diagrams for a bremsstrahlung emission of KK
    gravitons by electrons in the static electric field generated by
    atmospheric nuclei. \label{fig:2}}
\end{figure}

Nevertheless, herein we conjecture that KK graviton emission takes a
stand against processes driving the electromagnetic component of the
shower (viz. Bethe-Heitler pair production and electron
bremsstrahlung) rather than those generating the hadronic cascade; see
e.g. Fig.~\ref{fig:2}. Note that the number of electromagnetic
interactions surpasses the number of hadronic collisions in the shower by
orders of magnitude. Now, if KK production would only alter the
evolution of the electromagnetic component of the shower, then current simulations of extensive air
showers would underestimate the ratio of the {\it visible}
energy dissipated by the hadronic cascade $E_{\rm had}$ to that in the  
electromagnetic cascade $E_{\rm EM}$. Obviously, an underestimation of
the ratio 
$E_{\rm had}/E_{\rm EM}$ would alter the energy reconstruction of
the shower~\cite{Martynenko:2024rhj}. To accomodate the imbalance between $E_{\rm had}$ and
$E_{\rm EM}$, the cosmic
ray energy $E$ must be slightly
increased, and via (\ref{Nmu}) this implies an
increase in the number of muons that could account for the discrepancy
between simulations and experiment.

Note that in our conjecture the production of KK modes does not alter the
evolution of the hadronic shower. This is in agreement with Auger data
which indicate that the event-to-event fluctuations in the muon number
are well described by existing air shower
simulations~\cite{PierreAuger:2021qsd}.  Furthermore, a recent study  shows
that the distribution of atmospheric depth in which a cosmic ray
shower reaches its maximum number of particles (a.k.a. $X_{\rm max}$)
is systematically off in all interaction
models~\cite{PierreAuger:2024neu}. The best description of the data is
achieved if the model predictions are shifted to deeper $X_{\rm max}$
values. This new result is also in agreement with predictions of our
conjecture. While it is difficult to be
precise, if there were two micron-size dark dimensions gravity would
become strong around $10~{\rm TeV}$, an energy scale where discrepancies between model
building and cosmic-ray experimental data have
been observed. Ultimately, the impact of new scattering process on cosmic ray air 
showers, including those that spread out into a large internal space, can be best studied using a full-scale high-fidelity simulation. 

\section{Conclusions}
\label{sec:8}

The motivation for our work is the observation that two extra dimensions on the micron scale can simultaneously address the cosmological and gauge hierarchy problems. To be precise, the dark dimension scenario is not solving the problem of why there is a small cosmological constant, but it
 finds a model independent consequence, and if $n=2$ it is also connected to the electroweak hierarchy by low scale gravity. Consistency with astrophysical observations requires two extra
dimensions which do not admit isometries, whereby conservation of the
extra dimensional momentum is violated, allowing the massive KK modes
of the graviton to decay to other lighter graviton modes. Within this
scenario dark gravitons or else PBHs with masses in the range
$10^8 \alt M_{\rm BH}/{\rm g} \alt 10^{21}$ could make all
cosmological dark matter, though the PBH all dark-matter
interpretation requires some fine tuning. To this
conclusion, however, we must add an
important caveat: to remain consistent with cosmological observations
two extra dimensions of micron scale require a delicately fine tuning
of the temperature at which the universe enters the radiation
dominated epoch. Indeed, to overcome overproduction of KK modes at
early times the normalcy temperature is estimated to be $T_* \sim
2.15~{\rm MeV}$~\cite{Hall:1999mk}. We note that such a value of $T_*$
is comparable to the lowest possible
reheating temperature consistent with CMB data
and abundances predicted by BBN, which is estimated to be $T_* \sim 1.8~{\rm MeV}$~\cite{Hasegawa:2019jsa}.

Above and beyond searches of KK signals at future particle colliders, a
variety of astrophysical observations will be able to probe the ideas
discussed in this paper. For example, the asteroid-mass window for
PBHs to be all dark matter will be
tested in the near future by microlensing of X-ray
pulsars~\cite{Bai:2018bej,Tamta:2024pow}. This window 
include the interesting range ($10^{16.5} \alt M_{\rm BH}/{\rm g} \alt
10^{17.5}$) where an all dark matter interpretation in terms of 4D
Schwarzschild PBHs is excluded by the non-observation of their Hawking
radiation~\cite{Anchordoqui:2024tdj}. In addition, the long-baseline Deep Underground Neutrino Experiment
(DUNE) will probe relevant aspects of neutrino mixing in scenarios
with large extra dimensions~\cite{Berryman:2016szd,Siyeon:2024pte}. Finally,  AugerPrime full efficiency data taking (a.k.a. Auger Phase
II)  started in 2024 and is foreseen to add ten more years of
data~\cite{PierreAuger:2016qzd}. These data might help elucidate
whether KK production is at the center of the cosmic-ray {\it
  muon puzzle}. 

In conclusion, the dark dimension scenario does not provide a
theoretical prediction of why the cosmological constant is so small,
but it connects it to other scales, namely the size of the extra dimensions, and for the case of $n=2$ to TeV scale physics.
Actually, the 6D Planck scale in the
$10~{\rm TeV}$ regime is on the one hand already quite high in order
for being a ``natural'' solution of the electroweak hierarchy problem
and would already need some fine tuning. On the other hand, it is not
yet excluded by LHC data, however the next high luminosity LHC run
could possibly exclude it. As shown in our paper, the more severe
tension of the two extra dark dimension scenario comes from
cosmology. A further argument in favor of only one dark dimension
comes from gauge unification, which apparently needs a high species
scale~\cite{Heckman:2024trz}.  Yet another argument favoring $n=1$ comes from axion
  physics. On the one hand, for a single dark dimension, the QCD axion can be naturally localized on the SM
brane, where observational constraints force a narrow range for the
axion decay constant $10^9 \alt f_a/{\rm GeV} \alt 10^{10}$ while the axion mass,
$1 \alt m_a/{\rm meV} \alt 10$, is surprisingly near the mass scale
for the dark energy, the dark matter tower, and the
neutrino~\cite{Gendler:2024gdo}. On the other hand, for two dark
dimensions of micron size, a large value of $f_a$ consistent with observations~\cite{Hamaguchi:2018oqw} could come from the
suppression of the wave function if the axion lives in the
bulk, but
for $T_* \sim 2.15~{\rm MeV}$, such an axion would be severely constrained through its contribution to $N_{\rm eff}$.

\section*{Acknowledgements}

We like to thank Allen Caldwell, Ivano Basile, Gia Dvali, Georges Obied, Georg Raffelt, and Cumrun
Vafa for useful discussions. We thank Kevork Abazajian for permission
to reproduce Fig.~\ref{fig:1}. The work of L.A.A. is supported by the U.S. National Science
Foundation (NSF Grant PHY-2412679). I.A. is supported by the Second
Century Fund (C2F), Chulalongkorn University.  The work of D.L. is supported by the Origins
Excellence Cluster and by the German-Israel-Project (DIP) on
Holography and the Swampland. L.A.A. thanks the Theoretical Physics Laboratory at the UNLP in Argentina for hospitality during completion of this work.


\begin{thebibliography}{99}

\bibitem{Arkani-Hamed:1998jmv}
N.~Arkani-Hamed, S.~Dimopoulos and G.~R.~Dvali,
 {\color{rossoCP3}   The Hierarchy problem and new dimensions at a millimeter},
Phys. Lett. B \textbf{429}, 263-272 (1998)
doi:10.1016/S0370-2693(98)00466-3
[arXiv:hep-ph/9803315 [hep-ph]].

\bibitem{Antoniadis:1998ig}
I.~Antoniadis, N.~Arkani-Hamed, S.~Dimopoulos and G.~R.~Dvali,
{\color{rossoCP3} New dimensions at a millimeter to a Fermi and superstrings at a TeV},
Phys. Lett. B \textbf{436}, 257-263 (1998)
doi:10.1016/S0370-2693(98)00860-0
[arXiv:hep-ph/9804398 [hep-ph]].


\bibitem{Dvali:2007hz}
G.~Dvali,
{\color{rossoCP3} Black holes and large $N$ species solution to the hierarchy problem},
Fortsch. Phys. \textbf{58}, 528-536 (2010)
doi:10.1002/prop.201000009
[arXiv:0706.2050 [hep-th]].

\bibitem{Dvali:2007wp}
G.~Dvali and M.~Redi,
{\color{rossoCP3} Black hole bound on the number of species and quantum gravity at LHC},
Phys. Rev. D \textbf{77}, 045027 (2008)
doi:10.1103/PhysRevD.77.045027
[arXiv:0710.4344 [hep-th]].



\bibitem{Dvali:2009ks}
G.~Dvali and D.~L\"ust,
{\color{rossoCP3} Evaporation of Microscopic Black Holes in String Theory and the Bound on Species},
Fortsch. Phys. \textbf{58} (2010), 505-527
doi:10.1002/prop.201000008
[arXiv:0912.3167 [hep-th]].


\bibitem{Dvali:2010vm}
G.~Dvali and C.~Gomez,
{\color{rossoCP3} Species and Strings},
[arXiv:1004.3744 [hep-th]].


\bibitem{Dvali:2012uq}
G.~Dvali, C.~Gomez and D.~L\"ust,
{\color{rossoCP3} Black Hole Quantum Mechanics in the Presence of Species},
Fortsch. Phys. \textbf{61} (2013), 768-778
doi:10.1002/prop.201300002
[arXiv:1206.2365 [hep-th]].




\bibitem{Vafa:2005ui}
C.~Vafa,
{\color{rossoCP3}   The string landscape and the swampland},
[arXiv:hep-th/0509212 [hep-th]].




\bibitem{Palti:2019pca}
E.~Palti,
{\color{rossoCP3}   The swampland: introduction and review},
Fortsch. Phys. \textbf{67}, no.6, 1900037 (2019)
doi:10.1002/prop.201900037
[arXiv:1903.06239 [hep-th]].


\bibitem{vanBeest:2021lhn}
M.~van Beest, J.~Calder\'on-Infante, D.~Mirfendereski and I.~Valenzuela,
{\color{rossoCP3}   Lectures on the Swampland Program in string compactifications},
Phys. Rept. \textbf{989}, 1-50 (2022)
doi:10.1016/j.physrep.2022.09.002
[arXiv:2102.01111 [hep-th]].


\bibitem{Agmon:2022thq}
N.~B.~Agmon, A.~Bedroya, M.~J.~Kang and C.~Vafa,
{\color{rossoCP3}   Lectures on the string landscape and the swampland},
[arXiv:2212.06187 [hep-th]].

\bibitem{Ooguri:2006in}
H.~Ooguri and C.~Vafa,
{\color{rossoCP3}   On the Geometry of the String Landscape and the Swampland},
Nucl. Phys. B \textbf{766}, 21-33 (2007)
doi:10.1016/j.nuclphysb.2006.10.033
[arXiv:hep-th/0605264 [hep-th]].

\bibitem{Lust:2019zwm}
D.~L\"ust, E.~Palti and C.~Vafa,
 {\color{rossoCP3}   AdS and the Swampland},
Phys. Lett. B \textbf{797}, 134867 (2019)
doi:10.1016/j.physletb.2019.134867
[arXiv:1906.05225 [hep-th]].


\bibitem{Higuchi:1986py}
A.~Higuchi,
 {\color{rossoCP3} Forbidden mass range for spin-2 field theory in de Sitter space-time},
Nucl. Phys. B \textbf{282} (1987), 397-436
doi:10.1016/0550-3213(87)90691-2



\bibitem{Montero:2022prj}
M.~Montero, C.~Vafa and I.~Valenzuela,
{\color{rossoCP3} The dark dimension and the Swampland},
JHEP \textbf{02}, 022 (2023)
doi:10.1007/JHEP02(2023)022
[arXiv:2205.12293 [hep-th]].

\bibitem{Itoyama:1986ei}
H.~Itoyama and T.~R.~Taylor,
 {\color{rossoCP3} Supersymmetry restoration in the compactified $O(16) \times O(16)$-prime heterotic string theory},
Phys. Lett. B \textbf{186} (1987), 129-133
doi:10.1016/0370-2693(87)90267-X

\bibitem{Itoyama:1987rc}
H.~Itoyama and T.~R.~Taylor,
 {\color{rossoCP3} Small cosmological constant in string models},
FERMILAB-CONF-87-129-T.


\bibitem{Antoniadis:1991kh}
I.~Antoniadis and C.~Kounnas,
 {\color{rossoCP3} Superstring phase transition at high temperature},
Phys. Lett. B \textbf{261} (1991), 369-378
doi:10.1016/0370-2693(91)90442-S





\bibitem{Bonnefoy:2018tcp}
Q.~Bonnefoy, E.~Dudas and S.~L\"ust,
 {\color{rossoCP3}  On the weak gravity conjecture in string theory with broken supersymmetry},
Nucl. Phys. B \textbf{947} (2019), 114738
doi:10.1016/j.nuclphysb.2019.114738
[arXiv:1811.11199 [hep-th]].

\bibitem{Burgess:2023pnk}
C.~P.~Burgess and F.~Quevedo,
{\color{rossoCP3} Perils of towers in the swamp: dark dimensions and the robustness of EFTs},
JHEP \textbf{09} (2023), 159
doi:10.1007/JHEP09(2023)159
[arXiv:2304.03902 [hep-th]].


\bibitem{Branchina:2023ogv}
C.~Branchina, V.~Branchina, F.~Contino and A.~Pernace,
{\color{rossoCP3} Does the Cosmological Constant really indicate the existence of a Dark Dimension?},
doi:10.1142/S0219887824503055
[arXiv:2308.16548 [hep-th]].

\bibitem{Anchordoqui:2023laz}
L.~A.~Anchordoqui, I.~Antoniadis, D.~L\"ust and S.~L\"ust,
 {\color{rossoCP3} On the cosmological constant, the KK mass scale, and the cut-off dependence in the dark dimension scenario},
Eur. Phys. J. C \textbf{83} (2023) no.11, 1016
doi:10.1140/epjc/s10052-023-12206-2
[arXiv:2309.09330 [hep-th]].

\bibitem{Branchina:2024ljd}
C.~Branchina, V.~Branchina, F.~Contino and A.~Pernace,
{\color{rossoCP3} Dark Dimension and the Effective Field Theory limit},
doi:10.1142/S0219887824503031
[arXiv:2404.10068 [hep-th]].



\bibitem{Aoufia:2024awo}
C.~Aoufia, I.~Basile and G.~Leone,
{\color{rossoCP3} Species scale, worldsheet CFTs and emergent geometry},
JHEP \textbf{12} (2024), 111
doi:10.1007/JHEP12(2024)111
[arXiv:2405.03683 [hep-th]].




\bibitem{Basile:2024lcz}
I.~Basile and D.~L\"ust,
{\color{rossoCP3} Dark dimension with (little) strings attached},
[arXiv:2409.12231 [hep-th]].




\bibitem{Lee:2020zjt}
J.~G.~Lee, E.~G.~Adelberger, T.~S.~Cook, S.~M.~Fleischer and B.~R.~Heckel,
 {\color{rossoCP3}  New test of the gravitational $1/r^2$ law at separations down to 52 $\mu$m},
Phys. Rev. Lett. \textbf{124}, no.10, 101101 (2020)
doi:10.1103/PhysRevLett.124.101101
[arXiv:2002.11761 [hep-ex]].

\bibitem{Planck:2018vyg}
N.~Aghanim \textit{et al.} [Planck],
 {\color{rossoCP3}  Planck 2018 results  VI: Cosmological parameters},
Astron. Astrophys. \textbf{641}, A6 (2020)
[erratum: Astron. Astrophys. \textbf{652}, C4 (2021)]
doi:10.1051/0004-6361/201833910
[arXiv:1807.06209 [astro-ph.CO]].




\bibitem{FCC:2018vvp}
A.~Abada \textit{et al.} [FCC],
{\color{rossoCP3} FCC-hh: The Hadron Collider (Future Circular Collider Conceptual Design Report Volume 3)},
Eur. Phys. J. ST \textbf{228}, no.4, 755-1107 (2019)
doi:10.1140/epjst/e2019-900087-0




\bibitem{CMS:2012lmn}
S.~Chatrchyan \textit{et al.} [CMS],
{\color{rossoCP3}  Search for dark matter and large extra dimensions in $pp$ collisions yielding a photon and missing transverse energy},
Phys. Rev. Lett. \textbf{108}, 261803 (2012)
doi:10.1103/PhysRevLett.108.261803
[arXiv:1204.0821 [hep-ex]].


\bibitem{ATLAS:2012ezx}
G.~Aad \textit{et al.} [ATLAS],
{\color{rossoCP3}  Search for dark matter candidates and large extra dimensions in events with a photon and missing transverse momentum in $pp$ collision data at $\sqrt{s}=7$ TeV with the ATLAS detector},
Phys. Rev. Lett. \textbf{110}, no.1, 011802 (2013)
doi:10.1103/PhysRevLett.110.011802
[arXiv:1209.4625 [hep-ex]].

\bibitem{CMS:2011esc}
S.~Chatrchyan \textit{et al.} [CMS],
{\color{rossoCP3}  Search for new physics with a mono-jet and missing transverse energy in $pp$ collisions at $\sqrt{s} = 7$ TeV},
Phys. Rev. Lett. \textbf{107}, 201804 (2011)
doi:10.1103/PhysRevLett.107.201804
[arXiv:1106.4775 [hep-ex]].


\bibitem{ATLAS:2011kno}
G.~Aad \textit{et al.} [ATLAS],
{\color{rossoCP3}  Search for new phenomena with the monojet and missing transverse momentum signature using the ATLAS detector in $\sqrt{s}=7$ TeV proton-proton collisions},
Phys. Lett. B \textbf{705}, 294-312 (2011)
doi:10.1016/j.physletb.2011.10.006
[arXiv:1106.5327 [hep-ex]].



\bibitem{CMS:2014jvv}
V.~Khachatryan \textit{et al.} [CMS],
{\color{rossoCP3}  Search for dark matter, extra dimensions, and unparticles in monojet events in proton\textendash{}proton collisions at $\sqrt{s} = 8$ TeV},
Eur. Phys. J. C \textbf{75}, no.5, 235 (2015)
doi:10.1140/epjc/s10052-015-3451-4
[arXiv:1408.3583 [hep-ex]].

\bibitem{ATLAS:2015qlt}
G.~Aad \textit{et al.} [ATLAS],
{\color{rossoCP3} Search for new phenomena in final states with an energetic jet and large missing transverse momentum in pp collisions at $\sqrt{s}=$8 TeV with the ATLAS detector},
Eur. Phys. J. C \textbf{75}, no.7, 299 (2015)
[erratum: Eur. Phys. J. C \textbf{75}, no.9, 408 (2015)]
doi:10.1140/epjc/s10052-015-3517-3
[arXiv:1502.01518 [hep-ex]].

\bibitem{ATLAS:2021kxv}
G.~Aad \textit{et al.} [ATLAS],
{\color{rossoCP3}  Search for new phenomena in events with an energetic jet and missing transverse momentum in $pp$ collisions at $\sqrt {s}$ =13~TeV with the ATLAS detector},
Phys. Rev. D \textbf{103}, no.11, 112006 (2021)
doi:10.1103/PhysRevD.103.112006
[arXiv:2102.10874 [hep-ex]].


\bibitem{CMS:2021far}
A.~Tumasyan \textit{et al.} [CMS],
{\color{rossoCP3}  Search for new particles in events with energetic jets and large missing transverse momentum in proton-proton collisions at $ \sqrt{s} $ = 13 TeV},
JHEP \textbf{11}, 153 (2021)
doi:10.1007/JHEP11(2021)153
[arXiv:2107.13021 [hep-ex]].

\bibitem{D0:2008ayi}
V.~M.~Abazov \textit{et al.} [D0],
{\color{rossoCP3}  Search for large extra dimensions via single photon plus missing energy final states at $\sqrt{s}$ = 1.96-TeV},
Phys. Rev. Lett. \textbf{101}, 011601 (2008)
doi:10.1103/PhysRevLett.101.011601
[arXiv:0803.2137 [hep-ex]].

\bibitem{CDF:2008njt}
T.~Aaltonen \textit{et al.} [CDF],
{\color{rossoCP3}  Search for large extra dimensions in final states containing one photon or jet and large missing transverse energy produced in $p \bar{p}$ collisions at $\sqrt{s}$ = 1.96-TeV},
Phys. Rev. Lett. \textbf{101}, 181602 (2008)
doi:10.1103/PhysRevLett.101.181602
[arXiv:0807.3132 [hep-ex]].



\bibitem{CMS:2018ucw}
A.~M.~Sirunyan \textit{et al.} [CMS],
{\color{rossoCP3}  Search for new physics in dijet angular distributions using proton\textendash{}proton collisions at $\sqrt{s}=$ 13 TeV and constraints on dark matter and other models},
Eur. Phys. J. C \textbf{78}, no.9, 789 (2018)
[erratum: Eur. Phys. J. C \textbf{82}, no.4, 379 (2022)]
doi:10.1140/epjc/s10052-018-6242-x
[arXiv:1803.08030 [hep-ex]].


\bibitem{ATLAS:2017ayi}
M.~Aaboud \textit{et al.} [ATLAS],
{\color{rossoCP3}  Search for new phenomena in high-mass diphoton
  final states using 37 fb$^{-1}$ of proton--proton collisions collected at $\sqrt{s}=13$ TeV with the ATLAS detector},
Phys. Lett. B \textbf{775}, 105-125 (2017)
doi:10.1016/j.physletb.2017.10.039
[arXiv:1707.04147 [hep-ex]].


\bibitem{CMS:2018dqv}
A.~M.~Sirunyan \textit{et al.} [CMS],
{\color{rossoCP3}  Search for physics beyond the standard model in high-mass diphoton events from proton-proton collisions at $\sqrt{s} =$ 13 TeV},
Phys. Rev. D \textbf{98}, no.9, 092001 (2018)
doi:10.1103/PhysRevD.98.092001
[arXiv:1809.00327 [hep-ex]].


\bibitem{CMS:2021ctt}
A.~M.~Sirunyan \textit{et al.} [CMS],
{\color{rossoCP3}  Search for resonant and nonresonant new phenomena in high-mass dilepton final states at $ \sqrt{s} $ = 13 TeV},
JHEP \textbf{07}, 208 (2021)
doi:10.1007/JHEP07(2021)208
[arXiv:2103.02708 [hep-ex]].




\bibitem{Giudice:1998ck}
G.~F.~Giudice, R.~Rattazzi and J.~D.~Wells,
{\color{rossoCP3}  Quantum gravity and extra dimensions at high-energy colliders},
Nucl. Phys. B \textbf{544}, 3-38 (1999)
doi:10.1016/S0550-3213(99)00044-9
[arXiv:hep-ph/9811291 [hep-ph]].




\bibitem{Anchordoqui:2007da} 
  L.~A.~Anchordoqui, H.~Goldberg, S.~Nawata and T.~R.~Taylor,
{\color{rossoCP3}  Jet signals for low mass strings at the LHC,}
  Phys.\ Rev.\ Lett.\  {\bf 100}, 171603 (2008)
  doi:10.1103/PhysRevLett.100.171603
  [arXiv:0712.0386 [hep-ph]].



\bibitem{Anchordoqui:2008ac} 
  L.~A.~Anchordoqui, H.~Goldberg, S.~Nawata and T.~R.~Taylor,
  {\color{rossoCP3}  Direct photons as probes of low mass strings at the CERN LHC,}
  Phys.\ Rev.\ D {\bf 78}, 016005 (2008)
  doi:10.1103/PhysRevD.78.016005
  [arXiv:0804.2013 [hep-ph]].


\bibitem{Lust:2008qc}
D.~L\"ust, S.~Stieberger and T.~R.~Taylor,
{\color{rossoCP3}  The LHC string hunter's companion},
Nucl. Phys. B \textbf{808}, 1-52 (2009)
doi:10.1016/j.nuclphysb.2008.09.012
[arXiv:0807.3333 [hep-th]].


\bibitem{Anchordoqui:2008di} 
  L.~A.~Anchordoqui, H.~Goldberg, D.~L\"ust, S.~Nawata, S.~Stieberger and T.~R.~Taylor,
 {\color{rossoCP3}  Dijet signals for low mass strings at the LHC,}
  Phys.\ Rev.\ Lett.\  {\bf 101}, 241803 (2008)
  doi:10.1103/PhysRevLett.101.241803
  [arXiv:0808.0497 [hep-ph]].



\bibitem{Anchordoqui:2009mm} 
  L.~A.~Anchordoqui, H.~Goldberg, D.~L\"ust, S.~Nawata, S.~Stieberger and T.~R.~Taylor,
 {\color{rossoCP3}  LHC phenomenology for string hunters,}
  Nucl.\ Phys.\ B {\bf 821}, 181 (2009)
  doi:10.1016/j.nuclphysb.2009.06.023
  [arXiv:0904.3547 [hep-ph]].

\bibitem{Anchordoqui:2014wha} 
  L.~A.~Anchordoqui, I. Antoniadis, D-C. Dai, W.-Z. Feng,
  H. Goldberg, X. Huang, D. L\"ust, D. Stojkovic, and T.R. Taylor,
 {\color{rossoCP3}  String resonances at hadron colliders,}
  Phys.\ Rev.\ D {\bf 90}, no. 6, 066013 (2014)
  doi:10.1103/PhysRevD.90.066013
  [arXiv:1407.8120 [hep-ph]].


\bibitem{Khachatryan:2010jd} 
  V.~Khachatryan {\it et al.} [CMS],
{\color{rossoCP3}  Search for dijet resonances in 7~TeV $pp$ collisions at CMS},
  Phys.\ Rev.\ Lett.\  {\bf 105}, 211801 (2010)
  doi:10.1103/PhysRevLett.105.211801, 10.1103/PhysRevLett.106.029902
  [arXiv:1010.0203 [hep-ex]].




\bibitem{Chatrchyan:2011ns} 
  S.~Chatrchyan {\it et al.} [CMS],
 {\color{rossoCP3}  Search for resonances in the dijet mass spectrum from 7~TeV $pp$ collisions at CMS},
  Phys.\ Lett.\ B {\bf 704}, 123 (2011)
  doi:10.1016/j.physletb.2011.09.015
  [arXiv:1107.4771 [hep-ex]].



\bibitem{ATLAS:2011ai}
  G.~Aad {\it et al.} [ATLAS],
{\color{rossoCP3}  Search for production of resonant states in the photon-jet mass distribution using $pp$ collisions at $\sqrt{s}=7$ TeV collected by the ATLAS detector,}
  Phys.\ Rev.\ Lett.\  {\bf 108}, 211802 (2012)
  doi:10.1103/PhysRevLett.108.211802
  [arXiv:1112.3580 [hep-ex]].



\bibitem{Chatrchyan:2013qha} 
  S.~Chatrchyan {\it et al.} [CMS],
{\color{rossoCP3}  Search for narrow resonances using the dijet mass spectrum in $pp$ collisions at $\sqrt{s}$=8~TeV},
  Phys.\ Rev.\ D {\bf 87}, no. 11, 114015 (2013)
  doi:10.1103/PhysRevD.87.114015
  [arXiv:1302.4794 [hep-ex]].

 


\bibitem{Aad:2013cva} 
  G.~Aad {\it et al.} [ATLAS],
{\color{rossoCP3}  Search for new phenomena in photon+jet events collected in
    proton--proton collisions at $\sqrt{s} = 8~{\rm TeV}$ with the ATLAS detector},
  Phys.\ Lett.\ B {\bf 728}, 562 (2014)
  doi:10.1016/j.physletb.2013.12.029
  [arXiv:1309.3230 [hep-ex]].


\bibitem{Khachatryan:2015sja} 
  V.~Khachatryan {\it et al.} [CMS],
{\color{rossoCP3}  Search for resonances and quantum black holes using dijet mass spectra in proton-proton collisions at $\sqrt{s} =$ 8~TeV},
  Phys.\ Rev.\ D {\bf 91}, no. 5, 052009 (2015)
  doi:10.1103/PhysRevD.91.052009
  [arXiv:1501.04198 [hep-ex]].





\bibitem{Khachatryan:2015dcf}
V.~Khachatryan \textit{et al.} [CMS],
{\color{rossoCP3}  Search for narrow resonances decaying to dijets in proton-proton collisions at $\sqrt{s} =$ 13 TeV},
Phys. Rev. Lett. \textbf{116}, no.7, 071801 (2016)
doi:10.1103/PhysRevLett.116.071801
[arXiv:1512.01224 [hep-ex]].

  
 
\bibitem{Sirunyan:2016iap}
A.~M.~Sirunyan \textit{et al.} [CMS],
{\color{rossoCP3}  Search for dijet resonances in proton\textendash{}proton collisions at $\sqrt{s}$ = 13 TeV and constraints on dark matter and other models},
Phys. Lett. B \textbf{769}, 520-542 (2017)
[erratum: Phys. Lett. B \textbf{772}, 882-883 (2017)]
doi:10.1016/j.physletb.2017.02.012
[arXiv:1611.03568 [hep-ex]].

  


  
 
\bibitem{Sirunyan:2018xlo}
A.~M.~Sirunyan \textit{et al.} [CMS],
{\color{rossoCP3}  Search for narrow and broad dijet resonances in proton-proton collisions at $ \sqrt{s}=13 $ TeV and constraints on dark matter mediators and other new particles},
JHEP \textbf{08}, 130 (2018)
doi:10.1007/JHEP08(2018)130
[arXiv:1806.00843 [hep-ex]].




 
\bibitem{Sirunyan:2019vgj}
A.~M.~Sirunyan \textit{et al.} [CMS],
{\color{rossoCP3}  Search for high mass dijet resonances with a new background prediction method in proton-proton collisions at $\sqrt{s} =$ 13 TeV},
JHEP \textbf{05}, 033 (2020)
doi:10.1007/JHEP05(2020)033
[arXiv:1911.03947 [hep-ex]].



  


\bibitem{Arkani-Hamed:1998sfv}
N.~Arkani-Hamed, S.~Dimopoulos and G.~R.~Dvali,
{\color{rossoCP3}  Phenomenology, astrophysics and cosmology of theories with
submillimeter dimensions and TeV scale quantum gravity},
Phys. Rev. D \textbf{59}, 086004 (1999)
doi:10.1103/PhysRevD.59.086004
[arXiv:hep-ph/9807344 [hep-ph]].


\bibitem{Cullen:1999hc}
S.~Cullen and M.~Perelstein,
 {\color{rossoCP3} SN1987A constraints on large compact dimensions},
Phys. Rev. Lett. \textbf{83}, 268-271 (1999)
doi:10.1103/PhysRevLett.83.268
[arXiv:hep-ph/9903422 [hep-ph]].


\bibitem{Barger:1999jf}
V.~D.~Barger, T.~Han, C.~Kao and R.~J.~Zhang,
 {\color{rossoCP3} Astrophysical constraints on large extra dimensions},
Phys. Lett. B \textbf{461}, 34-42 (1999)
doi:10.1016/S0370-2693(99)00795-9
[arXiv:hep-ph/9905474 [hep-ph]].


\bibitem{Hanhart:2000er}
C.~Hanhart, D.~R.~Phillips, S.~Reddy and M.~J.~Savage,
 {\color{rossoCP3}  Extra dimensions, SN1987a, and nucleon-nucleon scattering data},
Nucl. Phys. B \textbf{595}, 335-359 (2001)
doi:10.1016/S0550-3213(00)00667-2
[arXiv:nucl-th/0007016 [nucl-th]].

\bibitem{Hanhart:2001fx}
C.~Hanhart, J.~A.~Pons, D.~R.~Phillips and S.~Reddy,
 {\color{rossoCP3} The likelihood of GODs' existence: Improving the SN1987a constraint on the size of large compact dimensions},
Phys. Lett. B \textbf{509}, 1-9 (2001)
doi:10.1016/S0370-2693(01)00544-5
[arXiv:astro-ph/0102063 [astro-ph]].


\bibitem{Hannestad:2003yd}
S.~Hannestad and G.~G.~Raffelt,
 {\color{rossoCP3} Supernova and neutron star limits on large extra dimensions reexamined},
Phys. Rev. D \textbf{67}, 125008 (2003)
[erratum: Phys. Rev. D \textbf{69}, 029901 (2004)]
doi:10.1103/PhysRevD.69.029901
[arXiv:hep-ph/0304029 [hep-ph]].

\bibitem{Hannestad:2001jv}
S.~Hannestad and G.~Raffelt,
 {\color{rossoCP3} New supernova limit on large extra dimensions},
Phys. Rev. Lett. \textbf{87}, 051301 (2001)
doi:10.1103/PhysRevLett.87.051301
[arXiv:hep-ph/0103201 [hep-ph]].


\bibitem{Han:1998sg}
T.~Han, J.~D.~Lykken and R.~J.~Zhang,
 {\color{rossoCP3}  On Kaluza-Klein states from large extra dimensions},
Phys. Rev. D \textbf{59}, 105006 (1999)
doi:10.1103/PhysRevD.59.105006
[arXiv:hep-ph/9811350 [hep-ph]].



\bibitem{Hall:1999mk}
L.~J.~Hall and D.~Tucker-Smith,
{\color{rossoCP3} Cosmological constraints on theories with large extra dimensions},
Phys. Rev. D \textbf{60}, 085008 (1999)
doi:10.1103/PhysRevD.60.085008
[arXiv:hep-ph/9904267 [hep-ph]].


\bibitem{Hannestad:2001xi}
S.~Hannestad and G.~G.~Raffelt,
{\color{rossoCP3}   Stringent neutron star limits on large extra dimensions},
Phys. Rev. Lett. \textbf{88}, 071301 (2002)
doi:10.1103/PhysRevLett.88.071301
[arXiv:hep-ph/0110067 [hep-ph]].





\bibitem{Mohapatra:2003ah}
R.~N.~Mohapatra, S.~Nussinov and A.~Perez-Lorenzana,
 {\color{rossoCP3}  Large extra dimensions and decaying $K K$ recurrences},
Phys. Rev. D \textbf{68}, 116001 (2003)
doi:10.1103/PhysRevD.68.116001
[arXiv:hep-ph/0308051 [hep-ph]].


\bibitem{Gonzalo:2022jac}
E.~Gonzalo, M.~Montero, G.~Obied and C.~Vafa,
{\color{rossoCP3} Dark Dimension Gravitons as Dark Matter},
JHEP \textbf{11}, 109 (2023)
doi:10.1007/JHEP11(2023)109
[arXiv:2209.09249 [hep-ph]].

\bibitem{Hannestad:2001nq}
S.~Hannestad,
 {\color{rossoCP3}  Strong constraint on large extra dimensions from cosmology},
Phys. Rev. D \textbf{64}, 023515 (2001)
doi:10.1103/PhysRevD.64.023515
[arXiv:hep-ph/0102290 [hep-ph]].

\bibitem{Fairbairn:2001ct}
M.~Fairbairn,
 {\color{rossoCP3}  Cosmological constraints on large extra dimensions},
Phys. Lett. B \textbf{508}, 335-339 (2001)
doi:10.1016/S0370-2693(01)00501-9
[arXiv:hep-ph/0101131 [hep-ph]].

\bibitem{Macesanu:2004gf}
C.~Macesanu and M.~Trodden,
 {\color{rossoCP3}  Relaxing cosmological constraints on large extra dimensions},
Phys. Rev. D \textbf{71}, 024008 (2005)
doi:10.1103/PhysRevD.71.024008
[arXiv:hep-ph/0407231 [hep-ph]].

\bibitem{Kawasaki:1999na}
M.~Kawasaki, K.~Kohri and N.~Sugiyama,
{\color{rossoCP3}  Cosmological constraints on late time entropy production},
Phys. Rev. Lett. \textbf{82}, 4168 (1999)
doi:10.1103/PhysRevLett.82.4168
[arXiv:astro-ph/9811437 [astro-ph]].



\bibitem{Kawasaki:2000en}
M.~Kawasaki, K.~Kohri and N.~Sugiyama,
{\color{rossoCP3} MeV scale reheating temperature and thermalization of neutrino background},
Phys. Rev. D \textbf{62}, 023506 (2000)
doi:10.1103/PhysRevD.62.023506
[arXiv:astro-ph/0002127 [astro-ph]].



\bibitem{Ichikawa:2005vw}
K.~Ichikawa, M.~Kawasaki and F.~Takahashi,
{\color{rossoCP3} The oscillation effects on thermalization of the neutrinos in the Universe with low reheating temperature},
Phys. Rev. D \textbf{72}, 043522 (2005)
doi:10.1103/PhysRevD.72.043522
[arXiv:astro-ph/0505395 [astro-ph]].


\bibitem{deSalas:2015glj}
P.~F.~de Salas, M.~Lattanzi, G.~Mangano, G.~Miele, S.~Pastor and O.~Pisanti,
 {\color{rossoCP3}  Bounds on very low reheating scenarios after Planck},
Phys. Rev. D \textbf{92}, no.12, 123534 (2015)
doi:10.1103/PhysRevD.92.123534
[arXiv:1511.00672 [astro-ph.CO]].

\bibitem{Hasegawa:2019jsa}
T.~Hasegawa, N.~Hiroshima, K.~Kohri, R.~S.~L.~Hansen, T.~Tram and S.~Hannestad,
{\color{rossoCP3} MeV-scale reheating temperature and thermalization of oscillating neutrinos by radiative and hadronic decays of massive particles},
JCAP \textbf{12}, 012 (2019)
doi:10.1088/1475-7516/2019/12/012
[arXiv:1908.10189 [hep-ph]].


\bibitem{Hannestad:2004px}
S.~Hannestad,
{\color{rossoCP3} What is the lowest possible reheating temperature?},
Phys. Rev. D \textbf{70}, 043506 (2004)
doi:10.1103/PhysRevD.70.043506
[arXiv:astro-ph/0403291 [astro-ph]].


\bibitem{Ichikawa:2006vm}
K.~Ichikawa, M.~Kawasaki and F.~Takahashi,
{\color{rossoCP3} Constraint on the effective number of neutrino species from the WMAP and SDSS LRG power spectra},
JCAP \textbf{05}, 007 (2007)
doi:10.1088/1475-7516/2007/05/007
[arXiv:astro-ph/0611784 [astro-ph]].


\bibitem{DeBernardis:2008zz}
F.~De Bernardis, L.~Pagano and A.~Melchiorri,
{\color{rossoCP3} New constraints on the reheating temperature of the universe after WMAP-5},
Astropart. Phys. \textbf{30}, 192-195 (2008)
doi:10.1016/j.astropartphys.2008.09.005

\bibitem{Barbieri:2025moq}
N.~Barbieri, T.~Brinckmann, S.~Gariazzo, M.~Lattanzi, S.~Pastor and O.~Pisanti,
{\color{rossoCP3} Current constraints on cosmological scenarios with very low reheating temperatures},
[arXiv:2501.01369 [astro-ph.CO]].



\bibitem{Abazajian:2023reo}
K.~N.~Abazajian and H.~G.~Escudero,
{\color{rossoCP3}   Visible in the laboratory and invisible in cosmology: Decaying sterile neutrinos},
Phys. Rev. D \textbf{108}, no.12, 123036 (2023)
doi:10.1103/PhysRevD.108.123036
[arXiv:2309.11492 [hep-ph]].

\bibitem{Escudero:2022rbq}
H.~G.~Escudero, J.~L.~Kuo, R.~E.~Keeley and K.~N.~Abazajian,
{\color{rossoCP3} Early or phantom dark energy, self-interacting, extra, or massive neutrinos, primordial magnetic fields, or a curved universe: An exploration of possible solutions to the $H_0$ and \ensuremath{\sigma}8 problems},
Phys. Rev. D \textbf{106}, no.10, 103517 (2022)
doi:10.1103/PhysRevD.106.103517
[arXiv:2208.14435 [astro-ph.CO]].





\bibitem{ParticleDataGroup:2016lqr}
C.~Patrignani \textit{et al.} [Particle Data Group],
{\color{rossoCP3} Review of Particle Physics},
Chin. Phys. C \textbf{40}, no.10, 100001 (2016)
doi:10.1088/1674-1137/40/10/100001

\bibitem{Friedrich:2020nze}
S.~Friedrich, G.~B.~Kim, C.~Bray, R.~Cantor, J.~Dilling, S.~Fretwell, J.~A.~Hall, A.~Lennarz, V.~Lordi and P.~Machule, \textit{et al.}
{\color{rossoCP3} Limits on the existence of sub-MeV sterile neutrinos from the decay of $^7$Be in superconducting quantum sensors},
Phys. Rev. Lett. \textbf{126}, no.2, 021803 (2021)
doi:10.1103/PhysRevLett.126.021803
[arXiv:2010.09603 [nucl-ex]].

\bibitem{Hu:1993gc}
W.~Hu and J.~Silk,
{\color{rossoCP3} Thermalization constraints and spectral distortions for massive unstable relic particles},
Phys. Rev. Lett. \textbf{70}, 2661-2664 (1993)
doi:10.1103/PhysRevLett.70.2661


\bibitem{Smith:2016vku}
P.~F.~Smith,
{\color{rossoCP3} Proposed experiments to detect keV range sterile neutrinos using energy-momentum reconstruction of beta decay or K-capture events},
New J. Phys. \textbf{21}, no.5, 053022 (2019)
doi:10.1088/1367-2630/ab1502
[arXiv:1607.06876 [physics.ins-det]].


\bibitem{KATRIN:2018oow}
S.~Mertens \textit{et al.} [KATRIN],
{\color{rossoCP3} A novel detector system for KATRIN to search for keV-scale sterile neutrinos},
J. Phys. G \textbf{46}, no.6, 065203 (2019)
doi:10.1088/1361-6471/ab12fe
[arXiv:1810.06711 [physics.ins-det]].

\bibitem{Lee:2025txi}
C.~Lee, X.~Zhang, A.~Kavner, T.~Parsons-Davis, D.~Lee, S.~T.~P.~Boyd, M.~Loidl, X.~Mougeot, M.~Rodrigues and M.~K.~Lee, \textit{et al.}
{\color{rossoCP3} Magneto-$\nu$: Heavy neutral lepton search using $^{241}$Pu $\beta^-$ decays},
[arXiv:2503.18350 [hep-ex]].

\bibitem{PTOLEMY:2019hkd}
M.~G.~Betti \textit{et al.} [PTOLEMY],
{\color{rossoCP3} Neutrino physics with the PTOLEMY project: active neutrino properties and the light sterile case},
JCAP \textbf{07}, 047 (2019)
doi:10.1088/1475-7516/2019/07/047
[arXiv:1902.05508 [astro-ph.CO]].

\bibitem{Dienes:2011ja}
K.~R.~Dienes and B.~Thomas,
{\color{rossoCP3} Dynamical Dark Matter: I. Theoretical Overview},
Phys. Rev. D \textbf{85}, 083523 (2012)
doi:10.1103/PhysRevD.85.083523
[arXiv:1106.4546 [hep-ph]].

\bibitem{Law-Smith:2023czn}
J.~A.~P.~Law-Smith, G.~Obied, A.~Prabhu and C.~Vafa,
{\color{rossoCP3}  Astrophysical constraints on decaying dark gravitons},
JHEP \textbf{06}, 047 (2024)
doi:10.1007/JHEP06(2024)047
[arXiv:2307.11048 [hep-ph]].


\bibitem{Peter:2010jy}
A.~H.~G.~Peter, C.~E.~Moody and M.~Kamionkowski,
{\color{rossoCP3} Dark-Matter Decays and Self-Gravitating Halos},
Phys. Rev. D \textbf{81}, 103501 (2010)
doi:10.1103/PhysRevD.81.103501
[arXiv:1003.0419 [astro-ph.CO]].

\bibitem{DES:2022doi}
S.~Mau \textit{et al.} [DES],
{\color{rossoCP3}  Milky Way satellite Census IV: Constraints on decaying dark matter from observations of Milky Way satellite galaxies},
Astrophys. J. \textbf{932}, no.2, 128 (2022)
doi:10.3847/1538-4357/ac6e65
[arXiv:2201.11740 [astro-ph.CO]].



\bibitem{Obied:2023clp}
G.~Obied, C.~Dvorkin, E.~Gonzalo and C.~Vafa,
{\color{rossoCP3}   Dark dimension and decaying dark matter gravitons},
Phys. Rev. D \textbf{109}, no.6, 063540 (2024)
doi:10.1103/PhysRevD.109.063540
[arXiv:2311.05318 [astro-ph.CO]].











\bibitem{Dienes:1998sb}
K.~R.~Dienes, E.~Dudas and T.~Gherghetta,
{\color{rossoCP3}  Neutrino oscillations without neutrino masses or heavy mass scales: A Higher dimensional seesaw mechanism},
Nucl. Phys. B \textbf{557}, 25 (1999)
doi:10.1016/S0550-3213(99)00377-6
[arXiv:hep-ph/9811428 [hep-ph]].

\bibitem{Arkani-Hamed:1998wuz}
N.~Arkani-Hamed, S.~Dimopoulos, G.~R.~Dvali and J.~March-Russell,
{\color{rossoCP3}  Neutrino masses from large extra dimensions},
Phys. Rev. D \textbf{65}, 024032 (2001)
doi:10.1103/PhysRevD.65.024032
[arXiv:hep-ph/9811448 [hep-ph]].

\bibitem{Dvali:1999cn}
G.~R.~Dvali and A.~Y.~Smirnov,
{\color{rossoCP3}  Probing large extra dimensions with neutrinos},
Nucl. Phys. B \textbf{563}, 63-81 (1999)
doi:10.1016/S0550-3213(99)00574-X
[arXiv:hep-ph/9904211 [hep-ph]].

\bibitem{Davoudiasl:2002fq}
H.~Davoudiasl, P.~Langacker and M.~Perelstein,
{\color{rossoCP3}  Constraints on large extra dimensions from neutrino oscillation experiments},
Phys. Rev. D \textbf{65}, 105015 (2002)
doi:10.1103/PhysRevD.65.105015
[arXiv:hep-ph/0201128 [hep-ph]].



\bibitem{Machado:2011jt}
P.~A.~N.~Machado, H.~Nunokawa and R.~Zukanovich Funchal,
{\color{rossoCP3}  Testing for large extra dimensions with neutrino oscillations},
Phys. Rev. D \textbf{84}, 013003 (2011)
doi:10.1103/PhysRevD.84.013003
[arXiv:1101.0003 [hep-ph]].


\bibitem{Forero:2022skg}
D.~V.~Forero, C.~Giunti, C.~A.~Ternes and O.~Tyagi,
{\color{rossoCP3}  Large extra dimensions and neutrino experiments},
Phys. Rev. D \textbf{106}, no.3, 035027 (2022)
doi:10.1103/PhysRevD.106.035027
[arXiv:2207.02790 [hep-ph]].


\bibitem{Carena:2017qhd}
M.~Carena, Y.~Y.~Li, C.~S.~Machado, P.~A.~N.~Machado and C.~E.~M.~Wagner,
{\color{rossoCP3}  Neutrinos in large extra dimensions and short-baseline $\nu_e$ appearance},
Phys. Rev. D \textbf{96}, no.9, 095014 (2017)
doi:10.1103/PhysRevD.96.095014
[arXiv:1708.09548 [hep-ph]].


\bibitem{Anchordoqui:2023wkm}
L.~A.~Anchordoqui, I.~Antoniadis and J.~Cunat,
{\color{rossoCP3}  Dark dimension and the standard model landscape},
Phys. Rev. D \textbf{109}, no.1, 016028 (2024)
doi:10.1103/PhysRevD.109.016028
[arXiv:2306.16491 [hep-ph]].

\bibitem{Burgess:2004yq}
C.~P.~Burgess, J.~Matias and F.~Quevedo,
{\color{rossoCP3}  MSLED: A Minimal supersymmetric large extra dimensions scenario},
Nucl. Phys. B \textbf{706}, 71-99 (2005)
doi:10.1016/j.nuclphysb.2004.11.025
[arXiv:hep-ph/0404135 [hep-ph]].


\bibitem{Anchordoqui:2024xvl}
L.~A.~Anchordoqui, I.~Antoniadis, D.~L\"ust and K.~Pe\~nal\'o~Castillo,
{\color{rossoCP3}  Cosmological constraints on dark neutrino towers},
Phys. Rev. D (in press)
[arXiv:2411.07029 [hep-ph]].

\bibitem{Zeldovich:1967lct}
Y.~B.~Zel'dovich and I.~D.~Novikov,
{\color{rossoCP3} The hypothesis of cores retarded during expansion and the hot cosmological model},
Soviet Astron. AJ (Engl. Transl. ), \textbf{10}, 602 (1967)


\bibitem{Hawking:1971ei}
S.~Hawking,
{\color{rossoCP3} Gravitationally collapsed objects of very low mass},
Mon. Not. Roy. Astron. Soc. \textbf{152}, 75 (1971)

\bibitem{Carr:1974nx}
B.~J.~Carr and S.~W.~Hawking,
{\color{rossoCP3} Black holes in the early Universe},
Mon. Not. Roy. Astron. Soc. \textbf{168}, 399-415 (1974)


\bibitem{Hodges:1990bf}
H.~M.~Hodges and G.~R.~Blumenthal,
{\color{rossoCP3}  Arbitrariness of inflationary fluctuation spectra},
Phys. Rev. D \textbf{42}, 3329-3333 (1990)
doi:10.1103/PhysRevD.42.3329


\bibitem{Carr:1993aq}
B.~J.~Carr and J.~E.~Lidsey,
{\color{rossoCP3}  Primordial black holes and generalized constraints on chaotic inflation},
Phys. Rev. D \textbf{48}, 543-553 (1993)
doi:10.1103/PhysRevD.48.543



\bibitem{Ivanov:1994pa}
P.~Ivanov, P.~Naselsky and I.~Novikov,
{\color{rossoCP3}  Inflation and primordial black holes as dark matter},
Phys. Rev. D \textbf{50}, 7173-7178 (1994)
doi:10.1103/PhysRevD.50.7173

\bibitem{Wu:2021gtd}
Y.~P.~Wu, E.~Pinetti and J.~Silk,
{\color{rossoCP3}  Cosmic coincidences of primordial-black-hole dark matter},
Phys. Rev. Lett. \textbf{128}, no.3, 031102 (2022)
doi:10.1103/PhysRevLett.128.031102
[arXiv:2109.09875 [astro-ph.CO]].

\bibitem{Leach:2001zf}
S.~M.~Leach, M.~Sasaki, D.~Wands and A.~R.~Liddle,
{\color{rossoCP3} Enhancement of superhorizon scale inflationary curvature perturbations},
Phys. Rev. D \textbf{64}, 023512 (2001)
doi:10.1103/PhysRevD.64.023512
[arXiv:astro-ph/0101406 [astro-ph]].


\bibitem{Affleck:1984fy}
I.~Affleck and M.~Dine,
{\color{rossoCP3} A new mechanism for baryogenesis},
Nucl. Phys. B \textbf{249}, 361-380 (1985)
doi:10.1016/0550-3213(85)90021-5

\bibitem{Hawking:1974rv}
S.~W.~Hawking,
 {\color{rossoCP3}  Black hole explosions},
Nature \textbf{248} (1974), 30-31
doi:10.1038/248030a0


\bibitem{Hawking:1975vcx}
S.~W.~Hawking,
 {\color{rossoCP3}  Particle creation by black holes},
Commun. Math. Phys. \textbf{43} (1975), 199-220
[erratum: Commun. Math. Phys. \textbf{46} (1976), 206]
doi:10.1007/BF02345020




\bibitem{Page:1976df}
D.~N.~Page,
 {\color{rossoCP3}  Particle emission rates from a black hole: Massless particles from an uncharged, nonrotating hole},
Phys. Rev. D \textbf{13} (1976), 198-206
doi:10.1103/PhysRevD.13.198


\bibitem{Carr:2020xqk}
B.~Carr and F.~Kuhnel,
{\color{rossoCP3} Primordial black holes as dark matter: Recent developments},
Ann. Rev. Nucl. Part. Sci. \textbf{70} (2020), 355-394
doi:10.1146/annurev-nucl-050520-125911
[arXiv:2006.02838 [astro-ph.CO]].


\bibitem{Green:2020jor}
A.~M.~Green and B.~J.~Kavanagh,
{\color{rossoCP3} Primordial Black Holes as a dark matter candidate},
J. Phys. G \textbf{48} (2021) no.4, 043001
doi:10.1088/1361-6471/abc534
[arXiv:2007.10722 [astro-ph.CO]].


\bibitem{Villanueva-Domingo:2021spv}
P.~Villanueva-Domingo, O.~Mena and S.~Palomares-Ruiz,
{\color{rossoCP3} A brief review on primordial black holes as dark matter},
Front. Astron. Space Sci. \textbf{8}, 87 (2021)
doi:10.3389/fspas.2021.681084
[arXiv:2103.12087 [astro-ph.CO]].

\bibitem{Argyres:1998qn}
P.~C.~Argyres, S.~Dimopoulos and J.~March-Russell,
{\color{rossoCP3} Black holes and submillimeter dimensions},
Phys. Lett. B \textbf{441} (1998), 96-104
doi:10.1016/S0370-2693(98)01184-8
[arXiv:hep-th/9808138 [hep-th]].


\bibitem{Anchordoqui:2022txe}
L.~A.~Anchordoqui, I.~Antoniadis and D.~L\"ust,
 {\color{rossoCP3} Dark dimension, the swampland, and the dark matter fraction composed of primordial black holes},
Phys. Rev. D \textbf{106}, no.8, 086001 (2022)
doi:10.1103/PhysRevD.106.086001
[arXiv:2206.07071 [hep-th]].

\bibitem{Anchordoqui:2022tgp}
L.~A.~Anchordoqui, I.~Antoniadis and D.~L\"ust,
 {\color{rossoCP3} The dark universe: Primordial black hole \ensuremath{\leftrightharpoons} dark graviton gas connection},
Phys. Lett. B \textbf{840}, 137844 (2023)
doi:10.1016/j.physletb.2023.137844
[arXiv:2210.02475 [hep-th]].

\bibitem{Anchordoqui:2024akj}
L.~A.~Anchordoqui, I.~Antoniadis and D.~L\"ust,
 {\color{rossoCP3}  The dark dimension, the swampland, and the dark matter fraction composed of primordial near-extremal black holes},
Phys. Rev. D \textbf{109}, no.9, 095008 (2024)
doi:10.1103/PhysRevD.109.095008
[arXiv:2401.09087 [hep-th]].

\bibitem{Anchordoqui:2024dxu}
L.~A.~Anchordoqui, I.~Antoniadis and D.~L\"ust,
{\color{rossoCP3} More on black holes perceiving the dark dimension},
Phys. Rev. D \textbf{110}, no.1, 015004 (2024)
doi:10.1103/PhysRevD.110.015004
[arXiv:2403.19604 [hep-th]].


\bibitem{Frolov:2002as}
V.~P.~Frolov and D.~Stojkovic,
{\color{rossoCP3} Black hole radiation in the brane world and recoil effect},
Phys. Rev. D \textbf{66}, 084002 (2002)
doi:10.1103/PhysRevD.66.084002
[arXiv:hep-th/0206046 [hep-th]].


\bibitem{Frolov:2002gf}
V.~P.~Frolov and D.~Stojkovic,
{\color{rossoCP3} Black hole as a point radiator and recoil effect in the brane world},
Phys. Rev. Lett. \textbf{89}, 151302 (2002)
doi:10.1103/PhysRevLett.89.151302
[arXiv:hep-th/0208102 [hep-th]].




\bibitem{Flachi:2005hi}
A.~Flachi and T.~Tanaka,
{\color{rossoCP3} Escape of black holes from the brane},
Phys. Rev. Lett. \textbf{95}, 161302 (2005)
doi:10.1103/PhysRevLett.95.161302
[arXiv:hep-th/0506145 [hep-th]].

\bibitem{Flachi:2006hw}
A.~Flachi, O.~Pujolas, M.~Sasaki and T.~Tanaka,
{\color{rossoCP3}  Black holes escaping from domain walls},
Phys. Rev. D \textbf{73}, 125017 (2006)
doi:10.1103/PhysRevD.73.125017
[arXiv:hep-th/0601174 [hep-th]].

\bibitem{Anchordoqui:2024jkn}
L.~A.~Anchordoqui, I.~Antoniadis, D.~L\"ust and K.~Pe\~nal\'o.~Castillo,
{\color{rossoCP3}  Bulk black hole dark matter},
Phys. Dark Univ. \textbf{46}, 101714 (2024)
doi:10.1016/j.dark.2024.101714
[arXiv:2407.21031 [hep-th]].


\bibitem{Tangherlini:1963bw}
F.~R.~Tangherlini,
{\color{rossoCP3} Schwarzschild field in $n$ dimensions and the dimensionality of space problem},
Nuovo Cim. \textbf{27} (1963), 636-651
doi:10.1007/BF02784569


\bibitem{Myers:1986un}
R.~C.~Myers and M.~J.~Perry,
 {\color{rossoCP3}  Black holes in higher dimensional space-times},
Annals Phys. \textbf{172} (1986), 304
doi:10.1016/0003-4916(86)90186-7

\bibitem{Anchordoqui:2001cg}
L.~A.~Anchordoqui, J.~L.~Feng, H.~Goldberg and A.~D.~Shapere,
 {\color{rossoCP3}  Black holes from cosmic rays: Probes of extra dimensions and new limits on TeV scale gravity},
Phys. Rev. D \textbf{65} (2002), 124027
doi:10.1103/PhysRevD.65.124027
[arXiv:hep-ph/0112247 [hep-ph]].

\bibitem{Anchordoqui:2018qom}
L.~A.~Anchordoqui,
 {\color{rossoCP3}  Ultra-high-energy cosmic rays},
Phys. Rept. \textbf{801}, 1-93 (2019)
doi:10.1016/j.physrep.2019.01.002
[arXiv:1807.09645 [astro-ph.HE]].

\bibitem{PierreAuger:2014ucz}
A.~Aab \textit{et al.} [Pierre Auger],
{\color{rossoCP3}  Muons in air showers at the Pierre Auger observatory: Mean number in highly inclined events},
Phys. Rev. D \textbf{91}, no.3, 032003 (2015)
[erratum: Phys. Rev. D \textbf{91}, no.5, 059901 (2015)]
doi:10.1103/PhysRevD.91.032003
[arXiv:1408.1421 [astro-ph.HE]].


\bibitem{PierreAuger:2016nfk}
A.~Aab \textit{et al.} [Pierre Auger],
 {\color{rossoCP3}  Testing hadronic interactions at ultrahigh energies with air showers measured by the Pierre Auger observatory},
Phys. Rev. Lett. \textbf{117}, no.19, 192001 (2016)
doi:10.1103/PhysRevLett.117.192001
[arXiv:1610.08509 [hep-ex]].


\bibitem{Albrecht:2021cxw}
J.~Albrecht, L.~Cazon, H.~Dembinski, A.~Fedynitch, K.~H.~Kampert, T.~Pierog, W.~Rhode, D.~Soldin, B.~Spaan and R.~Ulrich, \textit{et al.}
 {\color{rossoCP3}  The Muon Puzzle in cosmic-ray induced air showers and its connection to the Large Hadron Collider},
Astrophys. Space Sci. \textbf{367}, no.3, 27 (2022)
doi:10.1007/s10509-022-04054-5
[arXiv:2105.06148 [astro-ph.HE]].

\bibitem{Banks:1999gd}
T.~Banks and W.~Fischler,
 {\color{rossoCP3}   A model for high-energy scattering in quantum gravity},
[arXiv:hep-th/9906038 [hep-th]].

\bibitem{Anchordoqui:2003jr}
L.~A.~Anchordoqui, J.~L.~Feng, H.~Goldberg and A.~D.~Shapere,
 {\color{rossoCP3}  Updated limits on TeV scale gravity from absence of neutrino cosmic ray showers mediated by black holes},
Phys. Rev. D \textbf{68}, 104025 (2003)
doi:10.1103/PhysRevD.68.104025
[arXiv:hep-ph/0307228 [hep-ph]].

\bibitem{Cornet:2001gy}
F.~Cornet, J.~I.~Illana and M.~Masip,
 {\color{rossoCP3}  TeV strings and the neutrino nucleon cross-section at ultrahigh-energies},
Phys. Rev. Lett. \textbf{86}, 4235-4238 (2001)
doi:10.1103/PhysRevLett.86.4235
[arXiv:hep-ph/0102065 [hep-ph]].


\bibitem{Kazanas:2001ep}
D.~Kazanas and A.~Nicolaidis,
 {\color{rossoCP3}  Cosmic rays and large extra dimensions},
Gen. Rel. Grav. \textbf{35}, 1117-1123 (2003)
doi:10.1023/A:1024077103557
[arXiv:hep-ph/0109247 [hep-ph]].


\bibitem{Koch:2006yia}
B.~Koch, H.~J.~Drescher and M.~Bleicher,
  {\color{rossoCP3}  Gravitational radiation from ultrahigh energy cosmic rays in models with large extra dimensions},
Astropart. Phys. \textbf{25}, 291-297 (2006)
doi:10.1016/j.astropartphys.2006.02.007
[arXiv:astro-ph/0602164 [astro-ph]].



\bibitem{Martynenko:2024rhj}
N.~S.~Martynenko, G.~I.~Rubtsov, P.~S.~Satunin, A.~K.~Sharofeev and S.~V.~Troitsky,
 {\color{rossoCP3}  Hypothetical Lorentz invariance violation and the muon content of extensive air showers},
[arXiv:2412.08349 [astro-ph.HE]].


\bibitem{PierreAuger:2021qsd}
A.~Aab \textit{et al.} [Pierre Auger],
{\color{rossoCP3}  Measurement of the fluctuations in the number of
  muons in extensive air showers with the Pierre Auger Observatory},
Phys. Rev. Lett. \textbf{126}, no.15, 152002 (2021)
doi:10.1103/PhysRevLett.126.152002
[arXiv:2102.07797 [hep-ex]].


\bibitem{PierreAuger:2024neu}
A.~Abdul Halim \textit{et al.} [Pierre Auger],
{\color{rossoCP3}  Testing hadronic-model predictions of depth of maximum of air-shower profiles and ground-particle signals using hybrid data of the Pierre Auger Observatory},
Phys. Rev. D \textbf{109}, no.10, 102001 (2024)
doi:10.1103/PhysRevD.109.102001
[arXiv:2401.10740 [astro-ph.HE]].

\bibitem{Bai:2018bej}
Y.~Bai and N.~Orlofsky,
 {\color{rossoCP3}  Microlensing of X-ray pulsars: a method to detect primordial black hole dark matter},
Phys. Rev. D \textbf{99}, no.12, 123019 (2019)
doi:10.1103/PhysRevD.99.123019
[arXiv:1812.01427 [astro-ph.HE]].


\bibitem{Tamta:2024pow}
M.~Tamta, N.~Raj and P.~Sharma,
 {\color{rossoCP3}   Breaking into the window of primordial black hole dark matter with x-ray microlensing},
[arXiv:2405.20365 [astro-ph.HE]].

\bibitem{Anchordoqui:2024tdj}
L.~A.~Anchordoqui, I.~Antoniadis, D.~L\"ust and K.~Pe\~nal\'o~Castillo,
 {\color{rossoCP3}  Through the looking glass into the dark dimension: Searching for bulk black hole dark matter with microlensing of X-ray pulsars},
Phys. Dark Univ. \textbf{46}, 101681 (2024)
doi:10.1016/j.dark.2024.101681
[arXiv:2409.12904 [hep-ph]].


\bibitem{Berryman:2016szd}
J.~M.~Berryman, A.~de Gouv\^ea, K.~J.~Kelly, O.~L.~G.~Peres and Z.~Tabrizi,
 {\color{rossoCP3}  Large extra dimensions at the Deep Underground Neutrino Experiment},
Phys. Rev. D \textbf{94}, no.3, 033006 (2016)
doi:10.1103/PhysRevD.94.033006
[arXiv:1603.00018 [hep-ph]].


\bibitem{Siyeon:2024pte}
K.~Siyeon, S.~Kim, M.~Masud and J.~Park,
 {\color{rossoCP3}   Probing large extra dimension at DUNE using beam tunes},
JHEP \textbf{11}, 141 (2024)
doi:10.1007/JHEP11(2024)141
[arXiv:2409.08620 [hep-ph]].

\bibitem{PierreAuger:2016qzd}
A.~Aab \textit{et al.} [Pierre Auger],
 {\color{rossoCP3}  The Pierre Auger observatory upgrade:  Preliminary design report},
[arXiv:1604.03637 [astro-ph.IM]].

\bibitem{Heckman:2024trz}
J.~J.~Heckman, C.~Vafa, T.~Weigand and F.~Xu,
 {\color{rossoCP3}  Dark dimension and the grand unification of forces},
Phys. Rev. D \textbf{111} (2025) no.4, 046014
doi:10.1103/PhysRevD.111.046014
[arXiv:2409.01405 [hep-th]]. 








\bibitem{Gendler:2024gdo}
N.~Gendler and C.~Vafa,
{\color{rossoCP3}  Axions in the dark dimension},
JHEP \textbf{12}, 127 (2024)
doi:10.1007/JHEP12(2024)127
[arXiv:2404.15414 [hep-th]].


\bibitem{Hamaguchi:2018oqw}
K.~Hamaguchi, N.~Nagata, K.~Yanagi and J.~Zheng,
 {\color{rossoCP3}  Limit on the axion decay constant from the cooling neutron star in Cassiopeia A},
Phys. Rev. D \textbf{98}, no.10, 103015 (2018)
doi:10.1103/PhysRevD.98.103015
[arXiv:1806.07151 [hep-ph]].



\end{thebibliography}
\end{document}